\documentclass[aps,preprintnumbers,superscriptaddress,twocolumn,amsmath,amssymb,floatfix,pra]{revtex4} 
\pdfoutput=1
\usepackage{graphicx}
\usepackage[usenames,dvipsnames]{color}
\usepackage[colorlinks,bookmarks=false,citecolor=NavyBlue,linkcolor=OliveGreen,urlcolor=blue]{hyperref}
\newcommand\be{\begin{equation}}
\newcommand\bea{\begin{eqnarray}}
\newcommand\bes{\begin{subequations}}
\newcommand\esu{\end{subequations}}
\newcommand\ee{\end{equation}}
\newcommand\eea{\end{eqnarray}}
\newcommand\erf[1]        {\eqref{#1}}

\newcommand\w             {\omega}

\def\erf{\eqref}

\begin{document}

\title{Quantum Quench in a Harmonically Trapped One-Dimensional Bose Gas}

\author{Mario Collura}
\affiliation{The Rudolf Peierls Centre for Theoretical Physics, Oxford University, Oxford, OX1 3NP, United Kingdom.}

\author{M\'arton Kormos}
\affiliation{MTA-BME ``Momentum'' Statistical Field Theory Research Group, 1111 Budapest, Budafoki \'ut 8, Hungary}

\author{Pasquale Calabrese}
\affiliation{SISSA \& INFN, via Bonomea 265, 34136 Trieste, Italy}

\date{\today}

\begin{abstract}

We study the non-equilibrium dynamics of a one-dimensional Bose gas trapped by a harmonic potential for a quench from zero to infinite interaction. 
The different thermodynamic limits required for the equilibrium pre- and post-quench Hamiltonians are the origin of a few unexpected 
phenomena that have no counterparts in the translational invariant setting. 
We find that the dynamics is perfectly periodic with breathing time related to the strength of the trapping potential.  
For very short times,  we observe a sudden expansion leading to an extreme dilution of the gas and to the 
emergence of slowly decaying tails in the density profile. 
The haste of the expansion induces a undertow effect with a pronounced local minimum of the density at the center of the trap.
At half period there is a refocusing phenomenon characterized by a sharp central peak of the density, 
juxtaposed to algebraically decaying tails.
We finally show that the time-averaged density is correctly captured by a generalized Gibbs ensemble built with  
the conserved mode occupations.

\end{abstract}

\maketitle

\section{Introduction}

Over the last decade, the non-equilibrium dynamics of isolated quantum systems has been the focus of an intensive theoretical research. 
A main driving force behind this humongous theoretical work has been the constantly growing experimental activity on trapped ultracold 
atomic gases  which enabled the realization and the study of the non-equilibrium evolution of isolated quantum
systems \cite{kww-06,tetal-11,cetal-12,getal-11,shr-12,rsb-13,langen-2013,langen-2015,kaufman-2016}. 
The simplest non-equilibrium situation which attracted a lot of attention is the quantum quench \cite{cc-06}, 
i.e. the unitary time evolution from an initial state which is an eigenstate (usually the ground state) of a local Hamiltonian different from the 
one governing the evolution (for reviews see \cite{revq,gogolin-2015,calabrese-2016}).

An interesting outcome of both theoretical and experimental investigations is that generic and integrable systems show markedly different behavior. 
Generic systems relax to stationary states locally described by a thermal Gibbs ensemble with a temperature set by the conserved energy \cite{kaufman-2016,nonint,rdo-08,tvar,dalessio-2015}.
Conversely, integrable systems keep memory of the initial state also for infinite time because of the constraints imposed by
an infinite set of local and quasi-local conserved charges.
In this case, the stationary state is locally described by a generalized Gibbs ensemble in which all relevant conservation laws are taken into
account \cite{kww-06,langen-2015,gg,cazalilla-2006,barthel-2008,cramer-2008,cramer-2010,calabrese-2011,calabrese-2012,cic-12,fagotti-2013,p-13,fe-13b,sc-14,fcec-14,wdbf-14,PMWK14,KBC:Ising,ilievski-2015a,pvc-16,pvw-17,vidmar-2016,ef-16,rigol-16}.

However, most of the studies in the literature miss a very  important aspect of cold-atom experiments, namely, that experimental systems 
are not translationally invariant but the atoms are trapped by an external inhomogeneous (generally parabolic) potential. 
The presence of the trapping potential makes an exact analytic description very difficult  (if not impossible) in interacting many-body systems.

The interacting  one-dimensional Bose gas trapped by a harmonic potential is well described by the Hamiltonian 
\begin{multline}\label{HLL}
\hat H_c  =  \int_{-\infty}^{\infty} dx \bigg[\phi^{\dag} (x)\left(-\frac{1}{2}\partial^2_x +\frac{1}{2} \omega^2 x^2 \right) \hat\phi (x)\\
 +  c\, \hat\phi^{\dagger} (x) \hat\phi^{\dagger} (x) \hat\phi (x) \hat\phi(x)   \bigg]\,,
\end{multline}
where the boson field $\hat\phi(x)$ satisfies the canonical commutation relations
$[\hat\phi(x),\hat\phi^{\dag}(y)]=\delta(x-y)$, $c$ is the strength of the two-body interaction,
$\omega$ is the trap frequency, and we work in units such that $m=\hbar = 1$. 
In the absence of the external potential, the Hamiltonian \erf{HLL} is the
integrable Lieb--Liniger Hamiltonian \cite{LiebPR130} which is exactly solvable by means of Bethe ansatz 
for all values of the interaction strength $c$. 

Global interaction quenches correspond to the abrupt change of the coupling $c$ in the homogeneous system and were the focus of several works \cite{grd-10,fle-10,mc-12,ksc-13,nm-13,kcc14,nwbc-13,ds-13,ckc-14,cd-14,f-14,dc-14,dlc-15,bck-15,pe-16,zwkg-15,pce-16,fgkt-15,th-15,Bucc16,ccsh-16,bpc-16,bcs-17,npg-17}.
Other non-equilibrium situations of the Lieb-Liniger model, even in the presence of a trap, have also been studied 
\cite{mg-05,cro,ck-12,a-12,v-12,csc13,m-13,gn-14b,vwed-16,dsc-17,cgfb-14,carleo14,bencheikh17}, 
as well as the effects of a box confining potential \cite{mckc2014} and the consequences of different (anyonic) exchange statistics \cite{wrdk-14,pc-17}.
The solution of the quench problem for finite interaction strength $c$ in the post-quench Hamiltonian 
required the introduction of the quench action approach \cite{ce-13,caux-16} and the exact knowledge of the many-body overlaps
whose determination is a very difficult problem \cite{nwbc-13,kp-12,pozsgay-14,b-14,pc-14,lkz-15,hst-17,msca-16,ppv-17}. 
On the other hand,  for zero and infinite interaction more elementary techniques can be used to access the entire many-body dynamics.  
 
An external trapping potential breaks translational invariance and the Hamiltonian (\ref{HLL}) ceases to be integrable for generic couplings $c.$ 
There are two special values of $c$  where the model is still exactly solvable for arbitrary values of $\omega$.
For $c=0$ the Hamiltonian \erf{HLL} describes a free bosonic gas. 
Conversely, an infinite repulsion $c=+\infty$ makes the bosons impenetrable, and the gas can be mapped onto a system of free fermions \cite{TG}. For lattice models, this property was used to study the quench of a superlattice potential for hard-core bosons in the presence of a harmonic trap \cite{rigol-06}.

For the homogeneous case, analytic results were derived for the quench from $c=0$ to $c=\infty$ in Ref. \cite{kcc14}
(and generalized to other observables in \cite{dc-14,pc-17}). 
Despite the quadratic nature of the initial and final Hamiltonians, the non-equilibrium dynamics turned out to be highly non-trivial. In particular, Wick's theorem does not hold for finite times because the initial and final modes, one being bosonic and the other fermionic, are not linearly related. In order to mimic the experimental setups, the same quench was investigated in the presence of a hard wall confining potential in Ref. \cite{mckc2014}, where important differences were derived compared to the homogeneous case, e.g. the relaxation was found to take place in two steps.

In this paper we make a further step in closing the gap towards the description of current cold-atom experiments by studying the quench from 
the noninteracting to the strongly repulsive  gas confined by a {\it harmonic} potential. 
The study of this non-equilibrium dynamics suffers of all the complications already highlighted in homogeneous 
systems \cite{kcc14,mckc2014,pc-17}, in particular the absence of Wick's theorem at finite times due to the non-linear relation between 
pre- and post-quench modes. 
There is however a further and more cumbersome issue related to the nonexistence of a {\it unique} finite-density thermodynamic limit (TDL)
valid {\it both} for pre- and post-quench Hamiltonian.  
Indeed, for homogeneous systems the TDL is defined for arbitrary value of $c$ as the limit of the number of particles $N$ 
and the length of the system $L$ going to infinity with fixed density $N/L$. 
In a trapped system, the extension of the atomic cloud in the ground state is not fixed and depends on the interaction strength. 
In the case of free bosons ($c=0$), the ground state is the condensate in which all atoms are in the single-particle ground state of 
spatial extension $\ell^b\simeq \sqrt{1/\w}$. 
Conversely for $c=\infty$, the density of the effectively fermionic cloud is described by the famous Wigner semicircle law 
$n^{(f)}(x) = (2\sqrt{N\omega}/\pi)\sqrt{1/2- x^2 \omega/(4N)}$ of extension $\ell^f\simeq \sqrt{N/\w}$ \cite{xu-15}.
Thus in the free bosonic case, a finite density TDL is obtained taking $N,\ell^b\to\infty$ with $n=N/\ell^b\simeq N\sqrt\w$ kept constant, 
while in the impenetrable case  the correct limit is taken by keeping $n^f=N/\ell^f\simeq \sqrt {N\w}$ fixed \cite{rigol-06}.
The competition of these two different TDL's is one of the main difficulties in making the quench problem well-defined. 

To overcome this problem, all physical variables must be rescaled with the number of particles $N$ in non-standard way.
It is not straightforward to understand the correct rescaling because of the competition of the different length scales. 
Once the proper parameterizations have been understood, the limit of large $N$ can be taken and many of the computations become
almost elementary.
In this way we derive, among the other results, exact integral expressions for the time dependent density 
profile and show that its time average is entirely captured by a generalized Gibbs ensemble.

The paper is organized as follows. In Sec. \ref{sec:model} we review the Bose--Fermi mapping for the final Hamiltonian and introduce the 
technique to compute time evolved quantities. 
The basic building block is the initial {\em fermionic} correlation function which we determine analytically.
In Sec. \ref{sec:timedep} we derive analytic expressions for the time evolved fermionic two-point function and for the particle density. 
In Sec. \ref{sec:stat} we construct a generalized Gibbs ensemble in terms of the conserved mode occupations and show that it correctly 
describes the time averaged density and fermionic correlations. 
We give our conclusions in Sec. \ref{sec:concl}.

\section{Model and Quench}
\label{sec:model}

In this section we summarize some simple and well-known properties of pre- and post-quench Hamiltonians that are needed to study the 
quench dynamics of the coupling strength from $c=0$ to $c=\infty$  in a one-dimensional Bose gas trapped by a harmonic potential 
described by the Hamiltonian (\ref{HLL}).

\subsection{The initial setup}
The system is prepared in the $N$-particle ground state of the free-boson Hamiltonian $\hat H_0$, given by Eq. (\ref{HLL}) with $c=0.$
The quadratic Hamiltonian $\hat H_0$ can be diagonalized in terms of the one-particle creation and annihilation operators 
\be\label{xi_q}
\hat \xi_{q} =\int_{-\infty}^{\infty} dx \, \varphi^{*}_q(x) \, \hat\phi(x), 
\;\, \hat \xi^{\dag}_{q} = \int_{-\infty}^{\infty} dx \,  \varphi_q(x) \, \hat\phi^{\dag}(x)\,,
\ee
where the index $q$ is a non-negative integer and the one-particle eigenfunctions $\varphi_{q}(x)$
solve the Schr\"odinger equation of the 1D quantum harmonic oscillator
\be
\partial^2_x \varphi_{q}(x)/2 -\omega^2 x^2 \varphi_{q}(x)/2 = \epsilon_q \varphi_q(x)\,.
\ee
Explicitly, the normalized eigenfunctions are 
\be\label{eigenfunctions}
\varphi_q(x) =\frac{1}{\sqrt{2^q q!}} \left( \frac{\omega}{\pi} \right)^{1/4} 
 {\rm H}_{q}(x\sqrt{\omega}) \, {\rm e}^{-\omega x^2 /2},
\ee
where ${\rm H}_{q}(x) \equiv \partial^{q}_{s}\exp(2xs-s^2)|_{s=0}$ are the Hermite polynomials,
and $\epsilon_q = \omega(q+1/2)$ are the one-particle energy levels. 
The mode operators obey canonical commutation relations $[\hat \xi_{p},\hat \xi^\dag_{q}] = \delta_{p,q}.$

In terms of the modes $\hat \xi_q$ in Eq. (\ref{xi_q}), the pre-quench Hamiltonian $\hat H_0$ is diagonal:
\be\label{H_0_diag}
\hat H_0 = \sum_{q=0}^\infty \epsilon_q \, \hat \xi^{\dag}_q \hat\xi_q,
\ee
and the normalized $N$-particle ground state is the Bose-Einstein condensate 
\be\label{GS}
|\psi_0(N)\rangle=\frac1{\sqrt{N!}}(\hat\xi^{\dag}_{0})^{N}|0\rangle,
\ee
where $|0\rangle$ is the Fock vacuum (defined by $\hat\xi_q |0\rangle = 0,\; \forall q\in\mathbb{N}$).
The two-point correlation function of bosonic fields in the ground state can be computed 
using the relation $\langle \psi_0(N) | \hat\xi^{\dag}_p \hat\xi_q| \psi_0(N)\rangle = N \delta_{p,0}\delta_{q,0}$, 
yielding
\be
\langle \psi_0(N) | \hat\phi^{\dag}(x)\hat\phi(y)| \psi_0(N)\rangle 
=  \frac{n}{\sqrt{\pi}}  {\rm e}^{-\omega (x^2+y^2) /2},
\ee
where $n\equiv N\sqrt{\omega}$ is an average density given by the ratio of the total particle number and the oscillator length $1/\sqrt{\omega}.$ 
Setting $x=y$ we get the density profile
\be\label{eq_nx0}
n_{0}(x) \equiv \langle \psi_0(N) | \hat\phi^{\dag}(x)\hat\phi(x)| \psi_0(N)\rangle
 = \frac{n}{\sqrt{\pi}}  {\rm e}^{-\omega x^2}.
\ee

\subsection{The quench protocol}
 
The quench protocol considered in this paper consists of turning on an infinitely strong repulsion at $t=0$: 
the time evolution for $t>0$ is governed by $\hat H_\infty,$ the Hamiltonian (\ref{HLL}) with $c=\infty.$ 
This Hamiltonian describes the celebrated Tonks--Girardeau (TG) gas \cite{TG} which is a system of impenetrable bosons. 
To study this system, it is customary to  introduce hard-core bosonic fields $\hat\Phi(x)$, $\hat\Phi^{\dag}(x)$ which satisfy a hybrid algebra:
they obey an effective Pauli principle (induced by the infinite repulsion) at the same location but they commute at different points, 
\begin{equation}
[\hat{\Phi}(x),\hat{\Phi}^\dag(y)]=0,\,x\neq y, \quad[\hat{\Phi}^{\dag}(x)]^2=[\hat{\Phi}(x)]^2=0.
\label{alg2}
\end{equation} 
In terms of these hard-core fields, $\hat H_\infty$ is quadratic
\be\label{H_HCB}
\hat H_{\infty} = \int_{-\infty}^{\infty} dx \, \hat\Phi^{\dagger} (x) \left(-\frac{1}{2}\partial^2_x 
+\frac{1}{2}\omega^2 x^2 \right)  \hat\Phi (x),
\ee
and the infinitely strong repulsion is encoded in the hybrid commutation relations.

The hard-core boson fields can be mapped to fermionic fields \cite{TG} by the Jordan--Wigner transformation
\begin{subequations}
\label{JW}
\begin{align}
\hat{\Psi}(x) & =  \textnormal{exp}\left\{i\pi\int_0^xdz\hat{\Phi}^{\dag}(z)\hat{\Phi}(z)\right\}\hat{\Phi}(x), \\
\hat{\Psi}^{\dag}(x) & = \hat{\Phi}^{\dag}(x)\,\textnormal{exp}\left\{-i\pi\int_0^xdz\hat{\Phi}^{\dag}(z)\hat{\Phi}(z)\right\}.
\end{align}
\end{subequations}
The fermionic fields satisfy canonical anti-commutation relations $\{\hat\Psi(x),\hat\Psi^{\dag}(y)\}=\delta(x-y)$. 
The bosonic density operator $\hat\Phi^{\dag}(x)\hat\Phi(x)$ is mapped to the fermionic density operator $\hat\Psi^{\dag}(x)\hat\Psi(x).$
In terms of the fermionic fields the Hamiltonian (\ref{H_HCB}) becomes
\be\label{H_Fermi}
\hat H_{\infty} = \int_{-\infty}^{\infty} dx \, \hat\Psi^{\dagger} (x) \left(-\frac{1}{2}\partial^2_x 
+\frac{1}{2}\omega^2 x^2 \right) \hat\Psi (x),
\ee
that can be diagonalized by fermion mode operators $\hat\eta_{q}$, $\hat\eta^{\dag}_{q}$ defined by
\be\label{psi_eta}
\hat{\Psi}(x)=\sum_{q=0}^\infty \varphi_q(x) \hat\eta_q, \quad \hat\eta_q=\int_{-\infty}^{\infty} dx \, \varphi^{*}_q(x)\hat{\Psi}(x),
\ee
where the single-particle eigenfunctions $\varphi_q(x)$ are again given by Eq. (\ref{eigenfunctions}). 
In terms of the fermionic mode operators the Hamiltonian is diagonal
\begin{equation}
\label{Hwithnq}
\hat H_{\infty} = \sum_{q=0}^\infty \epsilon_q \,\hat\eta^{\dag}_q \hat\eta_q=
\sum_{q=0}^\infty \epsilon_q \, \hat n_{q},
\end{equation}
and we introduced  the fermionic mode occupation operators $\hat n_q \equiv \hat\eta^{\dag}_q \hat\eta_q$.

In the following we focus on the space and time dependent density as well as on the more general 
two-point {\em fermionic} correlation function
\be\label{fermi_corr_t}
C(x,y;t)\equiv 
\langle\exp(i \hat H t) \hat\Psi^{\dag}(x)\hat\Psi(y) \exp(-i \hat H t)\rangle,
\ee
where the notation $\langle\dots\rangle \equiv \langle\Psi_0(N)|\dots|\Psi_0(N)\rangle$ stands
for expectation values in the initial state. The density profile is given by the correlation function evaluated at $x=y,$ i.e. $n(x;t)= C(x,x;t).$

The time evolved correlation function can be expressed in terms of the post-quench modes as
\be\label{fermi_corr_t_eta}
C(x,y;t) = \sum_{p,q}\varphi^*_p(x)\varphi_q(y){\rm e}^{i(\epsilon_p-\epsilon_q)t}
\langle\hat\eta^{\dag}_p \hat\eta_q \rangle.
\ee
Another manageable expression for the correlator can be obtained by plugging (\ref{psi_eta}) into (\ref{fermi_corr_t_eta}), yielding
\be
\label{fermi_corr_t_green}
\begin{split}
C(x,y;t) =  \int_{-\infty}^{\infty}\!\!\!\! dx_0
\int_{-\infty}^{\infty} \!\!\!\! dy_0\,
&\mathcal{K}^{*}(x,x_0;t) \, \mathcal{K}(y,y_0;t) \\
&\times \langle \hat\Psi^{\dag}(x_0)\hat\Psi(y_0) \rangle,
\end{split}
\ee
where the kernel,
\be
\label{kernel}
\mathcal{K}(x,y;t) \equiv \sum_{q=0}^{\infty} \varphi_{q}(x)\varphi^{*}_{q}(y){\rm e}^{-i \epsilon_{q} t}
\ee
is the harmonic oscillator Green function, i.e.
\be
\langle x | \psi(t)\rangle = \int_{-\infty}^{\infty} \!\!dy \, \mathcal{K}(x,y;t) \langle y | \psi(0)\rangle.
\ee

Interestingly, the kernel has an analytical closed form obtainable using the mode functions \erf{eigenfunctions} and Mehler's formula \cite{mehler}
\be\label{mehler_formula}
\sum_{q=0}^{\infty} \frac{\rho^q}{2^q q!}H_{q}(x)H_{q}(y) 
= \frac{\exp \left[ - \frac{\rho^{2} (x^2 + y^2)-2\rho x y}{1-\rho^{2}} \right]}{\sqrt{1-\rho^2}},
\ee
leading to
\be
\mathcal{K}(x,y;t)   =  \left(\frac{\omega}{\pi}\right)^{1/2}
\frac{ \exp \left\{i\omega\frac{(x^2 + y^2)\cos(\omega t) - 2 x y  }{2\sin(\omega t)}  \right\}} 
{ \sqrt{2 i \sin(\omega t)} } .
\label{kernel1}
\ee
The kernel $\mathcal{K}(x,y;t) $ is a $2\pi$-periodic function in $\omega t,$ it is symmetric under the exchange of the 
space variables, $\mathcal{K}(y,x;t) = \mathcal{K}(x,y;t),$ and it satisfies the time-reversal
property $\mathcal{K}^{*}(x,y;t) = \mathcal{K}(x,y;-t).$
The kernel is not translational invariant since it depends both on $(x-y)$ and $(x+y)$. 

\begin{figure}[t!]
\includegraphics[width=0.45\textwidth]{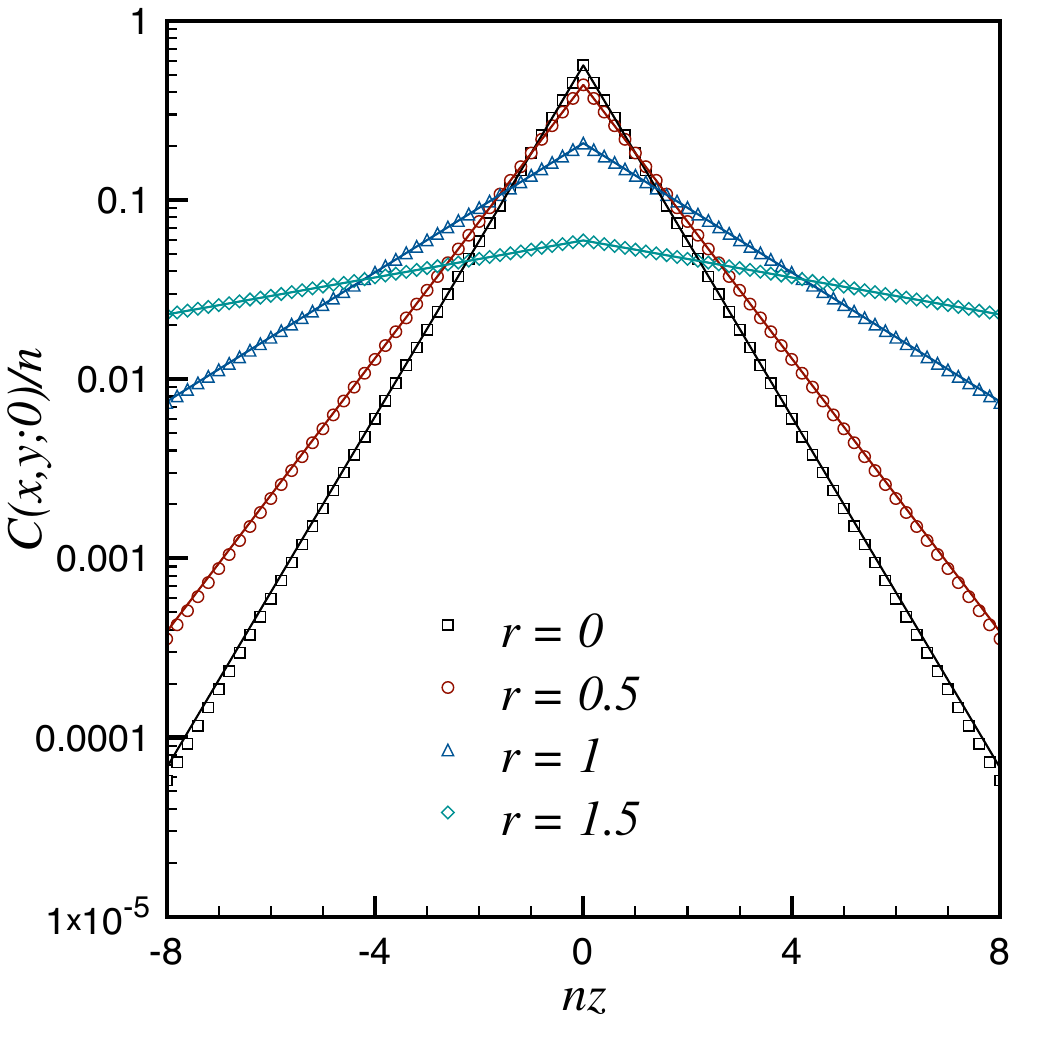}\\
\includegraphics[width=0.45\textwidth]{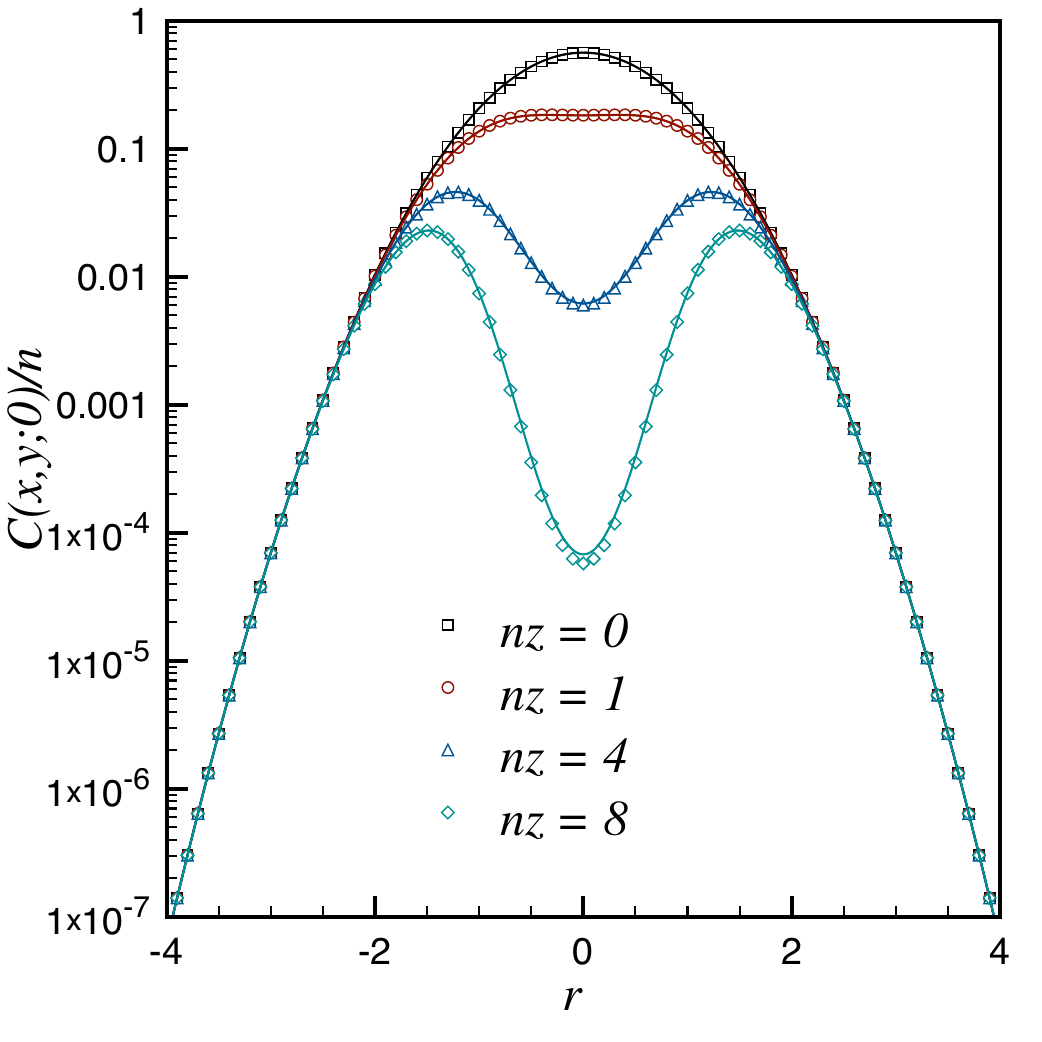}
\caption{Initial fermionic correlation function (\ref{PsiPsi_0_v2}) as a function of 
$nz=n(x-y)$ for fixed $r=\sqrt\w(x+y)/2$ (top panel) and as a function of $r$ for fixed $nz$ (bottom panel).
The numerical data for $N=1/\sqrt{\omega}= 200$ (symbols) are compared to the analytical scaling 
function given by Eq. (\ref{PsiPsi_0_TDL}) (solid lines). 
}
\label{fig_C00}
\end{figure}

\subsection{Initial fermionic correlation function}

The initial fermionic two-point function is a crucial ingredient for the calculation of the time evolved correlation function \erf{fermi_corr_t_green}.
Its calculation is not straightforward because of  
the non-Gaussian nature of the initial state in terms of the post-quench fermionic operators. 

The Jordan-Wigner transformation \erf{JW} relates the fermionic operators to the bosonic ones and can be used to rewrite the two-point function 
(for $x<y$) as \cite{kcc14,mckc2014}
\begin{multline}
\label{PsiPsi_0}
 \langle\hat\Psi^{\dag}(x)\hat\Psi(y)\rangle =
 \sum_{j=0}^{\infty}\frac{(-2)^j}{j!}\int_{x}^{y}dz_1\ldots\int_{x}^{y}dz_j \\
 \times 
 \langle\hat\Phi^{\dag}(x)\hat\Phi^{\dag}(z_1)\ldots\hat\Phi^{\dag}(z_j)\hat\Phi(z_j)\ldots \hat\Phi(z_1)\hat\Phi(y)\rangle,
 \end{multline}
 where the factor $(-2)^j$ results from normal ordering.
The multi-point correlator in \erf{PsiPsi_0} can be evaluated by replacing the hard-core boson fields with the canonical ones
(as justified in Refs. \cite{kcc14,mckc2014} from a lattice discretization), i.e.
\begin{multline}
\label{HCtoBoson}
\langle \hat\Phi^{\dag}(x)\hat\Phi^{\dag}(z_1)\ldots\hat\Phi^{\dag}(z_j)\hat\Phi(z_j)\ldots\hat\Phi(z_1)\hat\Phi(y)\rangle\\
=\langle \hat\phi^{\dag}(x)\hat\phi^{\dag}(z_1)\ldots\hat\phi^{\dag}(z_j)\hat\phi(z_j)\ldots \hat\phi(z_1)\hat\phi(y)\rangle.
 \end{multline}
The right hand side of Eq. (\ref{HCtoBoson}) is evaluated as a straightforward application of Wick's theorem:
\begin{multline}
\label{string}
\langle \hat\phi^{\dag}(x)\hat\phi^{\dag}(z_1)\ldots\hat\phi^{\dag}(z_j)\hat\phi(z_j)\ldots \hat\phi(z_1)\hat\phi(y)\rangle\\
=\varphi^{*}_0(x)\varphi_0(y) \, \prod_{i=1}^{j}|\varphi_0(z_i)|^2
 \langle (\hat \xi^{\dag}_0)^{ j+1} (\hat \xi_0)^{j+1}\rangle.
 \end{multline}
Using now $\hat\xi_{0} |\Psi_0(N)\rangle=\sqrt{N}|\Psi_0(N-1)\rangle$, 
we obtain $\langle (\hat \xi^{\dag}_0)^{ j+1} (\hat \xi_0)^{j+1}\rangle = N!/(N-j-1)!$ leading to
\begin{multline}
 \langle\hat\Psi^{\dag}(x)\hat\Psi(y)\rangle =   \varphi^{*}_0(x)\varphi_0(y)\\
 \times\sum_{j=0}^{\infty}\frac{(-2)^j}{j!}\frac{N!}{(N-j-1)!} 
\times \left|\int_x^y dz |\varphi_0(z)|^2\right|^j,
\end{multline}
where the absolute value takes into account the exchange of  integration limits for $x>y$.
Using 
\be
\left | \int_x^y dz |\varphi_0(z)|^2  \right | = \frac{1}{2}\left|{\rm Erf}(y\sqrt{\omega})-{\rm Erf}(x\sqrt{\omega})\right|
\ee
finally leads to
\begin{multline}
\label{PsiPsi_0_v2}
 \langle\hat\Psi^{\dag}(x)\hat\Psi(y)\rangle 
  =  \frac{N\sqrt{\omega}}{\sqrt{\pi}} {\rm e}^{-\omega(x^2+y^2)/2} \\
\times \left[1- \left|{\rm Erf}(y\sqrt{\omega})-{\rm Erf}(x\sqrt{\omega})\right| \right] ^{N-1}.
\end{multline}

Eq. (\ref{PsiPsi_0_v2}) is valid for any finite value of $\omega$ and $N$.
Let us discuss how we can obtain a consistent non-trivial result in the  thermodynamic limit (TDL).
In order to have a finite density we have to consider $N\to\infty$, $\omega \to 0$ with $N\sqrt{\omega}= n$.
At this point, we are forced to rescale distances with the trap frequency to avoid a trivial result. 
The proper rescaling leading to a non-trivial form is to keep $r\equiv\sqrt{\omega}(x+y)/2$ finite as $\omega\to 0$ while $z=x-y$ is finite
without rescaling.
In this regime, the limit of Eq. (\ref{PsiPsi_0_v2}) can be taken, leading to 
\be\label{PsiPsi_0_TDL}
 \langle\hat\Psi^{\dag}(x)\hat\Psi(y)\rangle 
 = \frac{n}{\sqrt{\pi}}\,{\rm e}^{-r^2} \exp \left( -\frac{2 n}{\sqrt{\pi}} {\rm e}^{-r^2} |z| \right). 
 \ee
The TDL defined in this way may seem rather artificial at first, but it is the only way to accommodate fermion correlations in a bosonic ground state. 
It is a direct consequence of the fact stressed in the introduction that bosonic and fermionic TDL's are different when the gas is trapped by a harmonic potential \cite{rigol-06}. 

In  Fig. \ref{fig_C00} we compare this asymptotic form with the exact finite $N$ expression (\ref{PsiPsi_0_v2}), reporting an excellent 
agreement in the TDL.
In what follows, Eq. (\ref{PsiPsi_0_TDL}) is the starting point 
to analytically compute the time evolved two-point function (\ref{fermi_corr_t_green}) and the fermionic mode occupations.

 \begin{figure}[t!]
\includegraphics[width=0.45\textwidth]{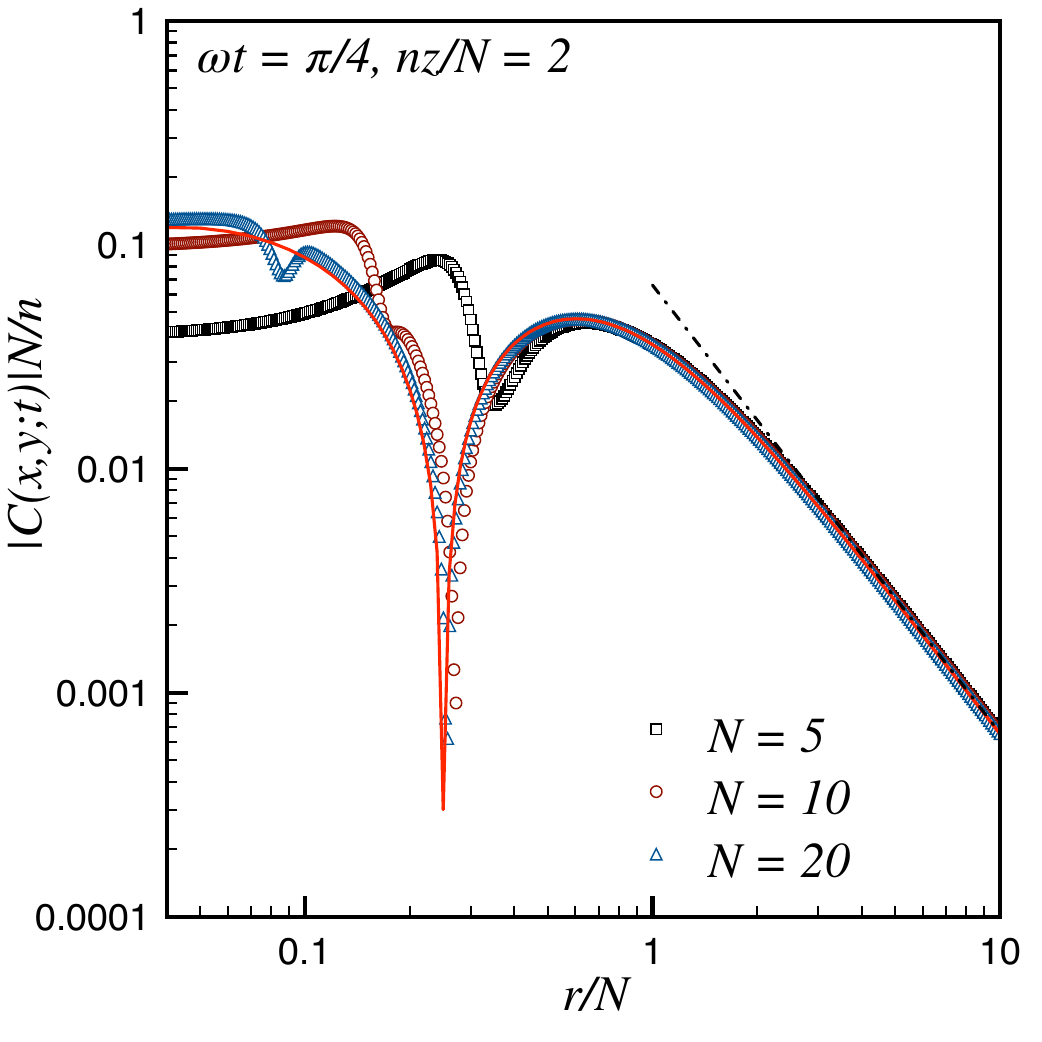}\\
\includegraphics[width=0.45\textwidth]{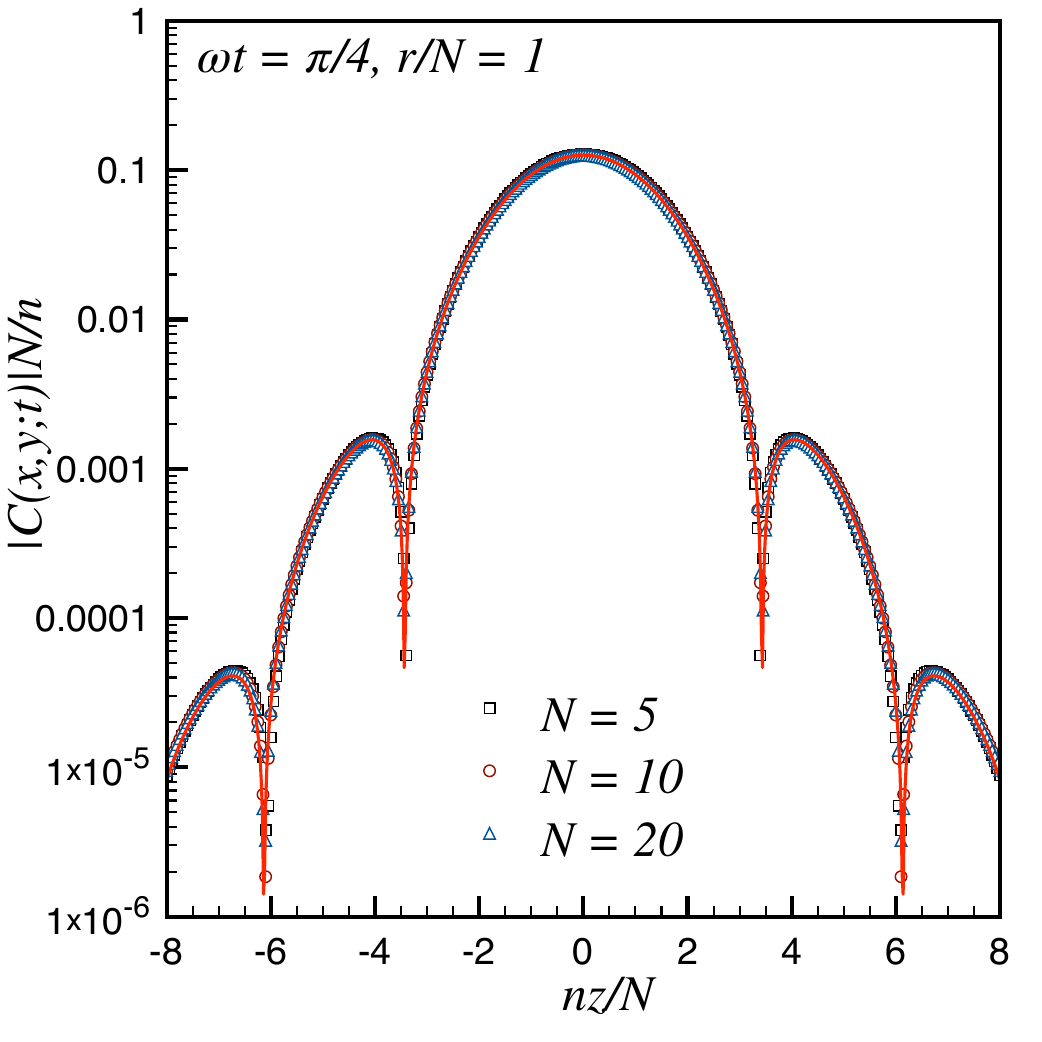}\\
\caption{Fermionic correlation function at fixed time $\w t=\pi/4$ as a function of the distance $r$ from the center of the trap 
at fixed separation $z$ (top panel) and the other way around (bottom panel). 
The symbols are the exact data at finite $N$ (the apparent singularities are just sign changes in logarithmic scale). 
The full red line in the top (bottom) panel is the large $r$ asymptotics in Eq. \erf{Casympt1} (Eq. \erf{Casympt2}), where the first twenty terms in the sum have been kept. 
The dot-dashed line in the left panel is the very large $r$ algebraic tail, i.e. the first term of Eq. \erf{Casympt2}.
} 
\label{fig_C}
\end{figure}

\section{Time-dependent quantities}\label{sec:timedep}

In this section we explicitly compute the time evolution of  the two-point fermionic correlation function and of the density profile.
Due to the periodicity of the kernel \erf{kernel1}, we do not expect relaxation for large times but oscillatory behavior. 

\subsection{Fermionic correlation function}
The starting point of our calculation is to apply the integral representation of the correlation function (\ref{fermi_corr_t_green}) 
using the explicit form of the harmonic oscillator kernel \erf{kernel1} and the the initial two-point function (\ref{PsiPsi_0_TDL}).
The two-point fermionic correlation function evolves under the simultaneous action of 
two harmonic kernel operators (c.f. Eq. \erf{fermi_corr_t_green}). The product of the kernels can be written as
\begin{multline}
\label{KK}
\mathcal{K}^{*}(x,x_0;t)\mathcal{K}(y,y_0;t) \\
= \omega\frac{\exp\left\{i\sqrt\omega \frac{z_0(r-r_0\cos(\omega t))+z(r0-r\cos(\omega t))}{\sin(\omega t)}\right\}}{2\pi|\sin(\omega t)|},
\end{multline}
where, following the recipe for the TDL introduced at the end of the previous section, we introduced the variables
\be
\begin{aligned}
r&=\sqrt\w\,\frac{x+y}{2},& z&=x-y,\\
r_0&=\sqrt\w\,\frac{x_0+y_0}{2},& z_0&=x_0-y_0.
\end{aligned}
\ee
Changing integration variables in Eq. (\ref{fermi_corr_t_green}) from $(x_0,y_0)$ to $(r_0,z_0)$, 
the $z_0$-integral can be evaluated analytically 
using $A \int_{-\infty}^{\infty} dz_0 {\rm e}^{-2 A |z_0|} {\rm e}^{i \alpha z_0} = 1/[1+\alpha^2/(4A^2)]$. 
Thus the correlation function can be written as a single integral as 
\begin{multline}
\label{PsiPsi_t_TDL}
C(x,y;t)  \\=  \frac{\sqrt{\omega}}{2\pi|\sin(\w t)|}
 \int_{-\infty}^{\infty}dr_0
\frac{\exp \left[\frac{i\sqrt{\omega}}{\sin(\w t)}z(r_0- r\cos(\w t))\right]} 
{1+\left(\frac{ \sqrt{\pi}(r- r_0\cos(\w t))}{2N \sin(\omega t)}{\rm e}^{r_0^2}\right)^2} .
\end{multline}
It is cumbersome, but straightforward to show that as $t\to0$ we recover the initial correlation function \erf{PsiPsi_0_TDL}. 
The correlator is periodic with period $T=\pi/\w.$
As a fundamental test, we checked numerically that evolving the finite size correlation  \erf{PsiPsi_0_v2}  with the 
kernel \erf{fermi_corr_t_green} we recover Eq. \erf{PsiPsi_t_TDL} for large $N$.

 \begin{figure*}[t!]
\includegraphics[width=0.96\textwidth]{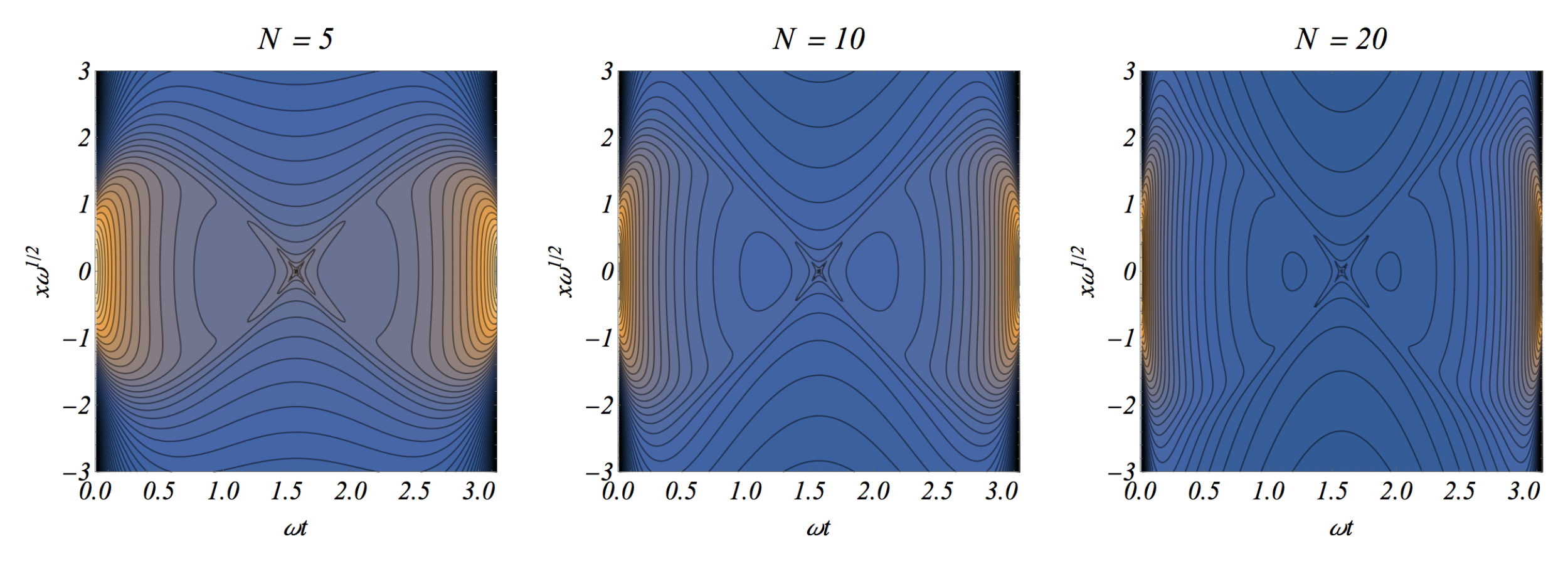}\\
\caption{Contour plots of the rescaled density $n(x;t)/n$ given by Eq. (\ref{eq_nxt}) for different number of particles $N.$ 
} 
\label{fig_nxt_color}
\end{figure*}

The correlator in Eq. \erf{PsiPsi_t_TDL} is well defined everywhere except when $\omega t = \pi/2$ and $r = 0$ simultaneously
(i.e. for two points symmetric with respect to the center of the trap at half period). 
In this case the integral (\ref{PsiPsi_t_TDL}) diverges because the integrand  is $e^{i \sqrt{\pi/2} z r_0}$.
While the integral  genuinely diverges in the TDL, it is important to work out the $N$-dependence of this divergence.
In order to do so, we first identify the origin of the divergence with the fact that replacing Eq. \erf{PsiPsi_0_v2} with Eq. \erf{PsiPsi_0_TDL} 
we dropped a subleading Gaussian factor $\sim \exp(-\w z^2 /4)$ which becomes important {\it only} for $\w t=\pi/2$ and $r=0$ 
when it induces an $N$-dependent cutoff for the $r_0$-integral. 
Reintroducing this factor, the $z_0$-integral becomes 
$$n_0\int dz_0 e^{-2n_0|z_0|}e^{-\w z_0^2/4}=2n_0 \sqrt{\frac\pi\w} e^{4 n_0^2 /\w} {\rm erfc}\frac{2n_0}{\sqrt{\w}},$$ 
where $n_0\equiv n_0(r_0)=n/\sqrt\pi e^{-r_0^2}$  is the initial density \erf{eq_nx0} in $r_0$. 
Taking now the TDL $N\to\infty$ with $n$ constant,  this function of $r_0$ approaches a box function of height $1.$ 
The edges of the box are located at the inflection points $\pm r_0^*$
which numerically are very well approximated by $r_0^*\approx \sqrt{\ln(4N/\sqrt\pi)}$. 
Then, at this very special point the correlation function is 
\be
C\Big(x,-x;\frac{\pi}{2\w}\Big) \approx \frac{\sqrt{\omega}}{2\pi} \int_{-r_0^*}^{r_0^*}dr_0=
\frac{\sqrt{\w}}\pi \sqrt{\ln (4N/\sqrt\pi)}.
\ee


There is another interesting limit of the correlation function that can be worked out more explicitly. 
Indeed, the integral in Eq. \erf{PsiPsi_t_TDL} can be simplified when the two points are far from the center of the trap. 
In this limit, the gaussian factor in the denominator of the integrand cuts off the integral around $r_0\approx r^* = \sqrt{\ln(2N\sin\w t/\sqrt\pi)}.$ 
Thus for $r\gg r^*$ the term $r_0\cos(\w t)$ can be neglected. 
Under this assumption, the integral can be performed by expanding  $e^{i \sqrt\w z r_0/\sin(\w t)}$ in powers of $z$, obtaining 
\begin{multline}
\label{Casympt1}
C(x,y;t)  \approx  - \frac{\sqrt\w \, {\rm e}^{-i \zeta  r \cos(\w t)}}{2\sqrt{2\pi}|\sin(\w t)|} \\
\times \sum_{k=0}^{\infty} \frac{(-\zeta^2 /8)^k}{k!}
{\rm Li}_{k+1/2}\left(-\frac{1}{\rho^2} \right)\, , 
\end{multline}
where we introduced the new variables $\zeta = \sqrt\w z /\sin(\w t)$ and $\rho = \sqrt\pi r /[2N \sin(\w t)]$.
Here ${\rm Li}_{s}(z) \equiv \sum_{j=1}^{\infty}z^j/(j^s)$ is the polylogarithm function.
Using the series representation of the polylogarithm function we finally obtain
\be
\label{Casympt2}
C(x,y;t) \approx  - \frac{\sqrt\w  {\rm e}^{-i \zeta  r \cos(\w t)}}{2\sqrt{2\pi}|\sin(\w t)|}
\sum_{j=1}^{\infty}\frac{{\rm e}^{-\frac{\zeta^2}{8 j}}}{\sqrt{j}} \left(-\frac{1}{\rho^2} \right)^{j} \, .
\ee
When $r$ is very large, specifically for $\rho\gg1$ (i.e.  $ r \gg 2N \sin(\w t)/\sqrt\pi$), the leading term in the sum \erf{Casympt2} 
is the one with $j=1$, and hence the correlation function decays as $r^{-2}$ with the distance from the center of the trap. 
In this regime, the two-point function decays as a Gaussian of the separation $z$ with a typical length $\sim \sin\w t/\sqrt\w$.
However, the sum in Eq. \erf{Casympt2} (or equivalently \erf{Casympt1}) describes the entire regime $r\gg r^*$.
Consequently, for large enough $N$, there is always an intermediate window of $r$, i.e.
$\sqrt{\ln(2N\sin\w t/\sqrt\pi)}\alt r\alt 2N \sin(\w t)/\sqrt\pi$, in which several terms in the sum \erf{Casympt2} are needed to capture 
the correct behavior. 

In order to show the rough behavior of the fermionic correlation function, in Fig. \ref{fig_C} we report two selected plots for 
$\omega t=\pi/4$. 
Both plots show how Eq. \erf{Casympt1} describes very accurately the correlation function for moderately large values of $r$ and $N$.
In the left panel, it is also shown how the asymptotic power law tail for large $r$ in Eq. \erf{Casympt2} sets up at $r\sim O(N)$.
We carefully checked that a similar agreement is found for arbitrary values of $\omega t$.

 \begin{figure*}[t!]
\includegraphics[width=0.33\textwidth]{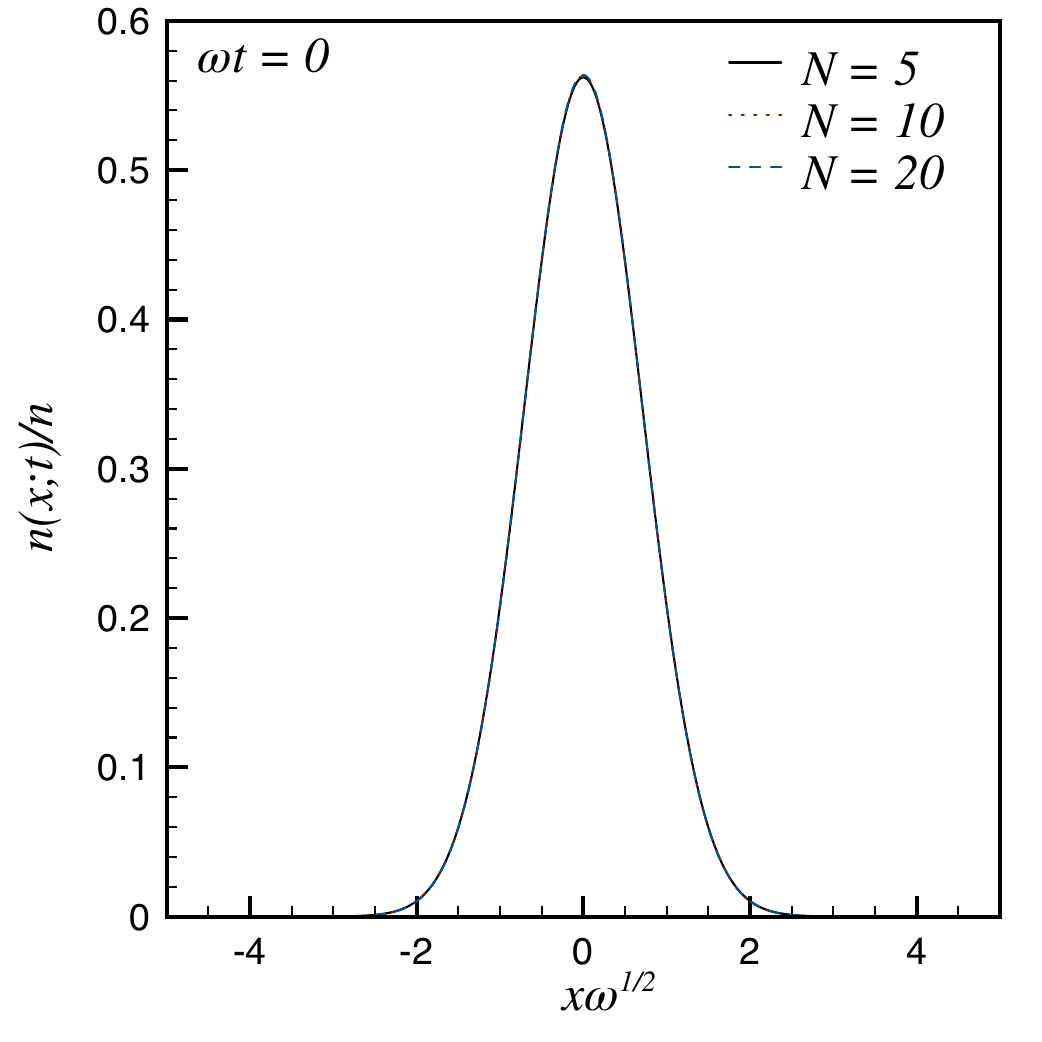}\includegraphics[width=0.33\textwidth]{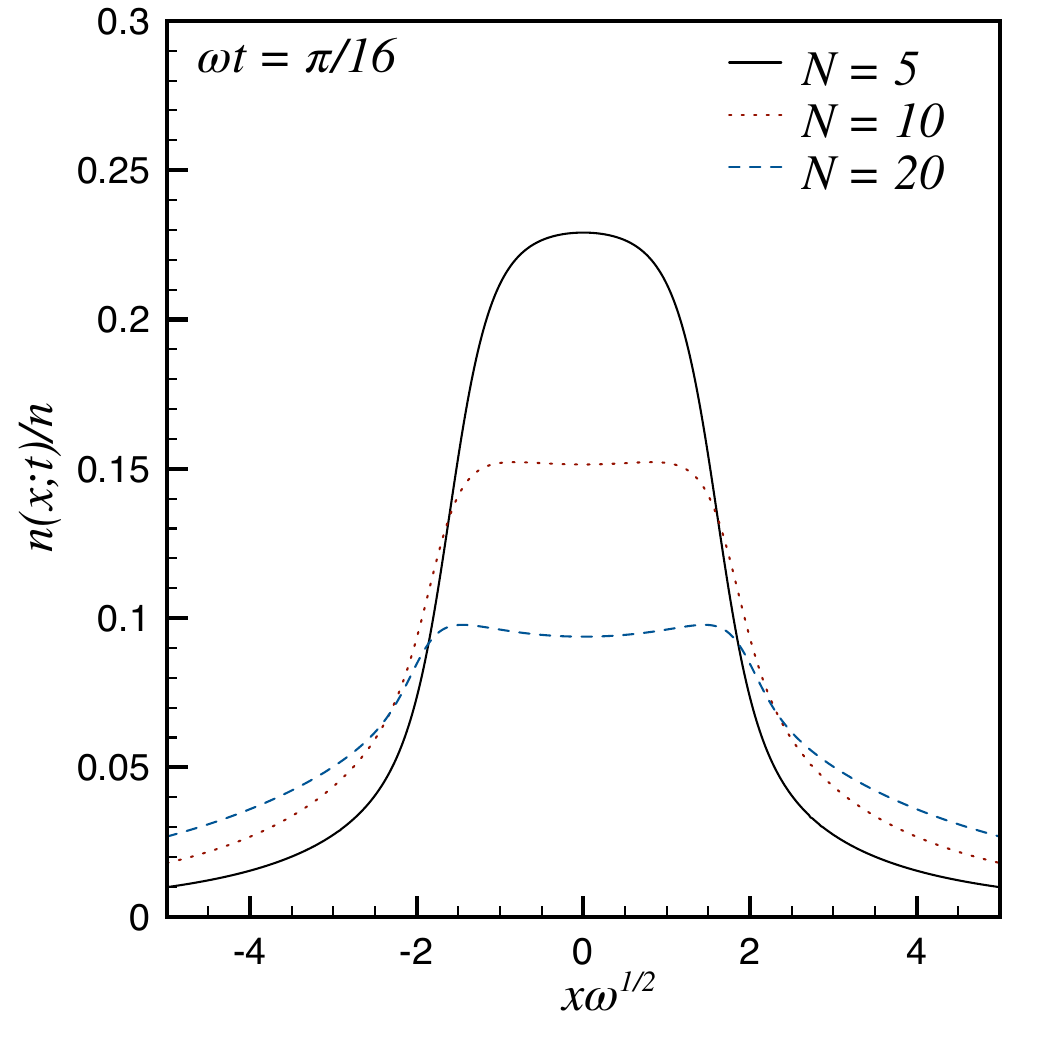}\includegraphics[width=0.33\textwidth]{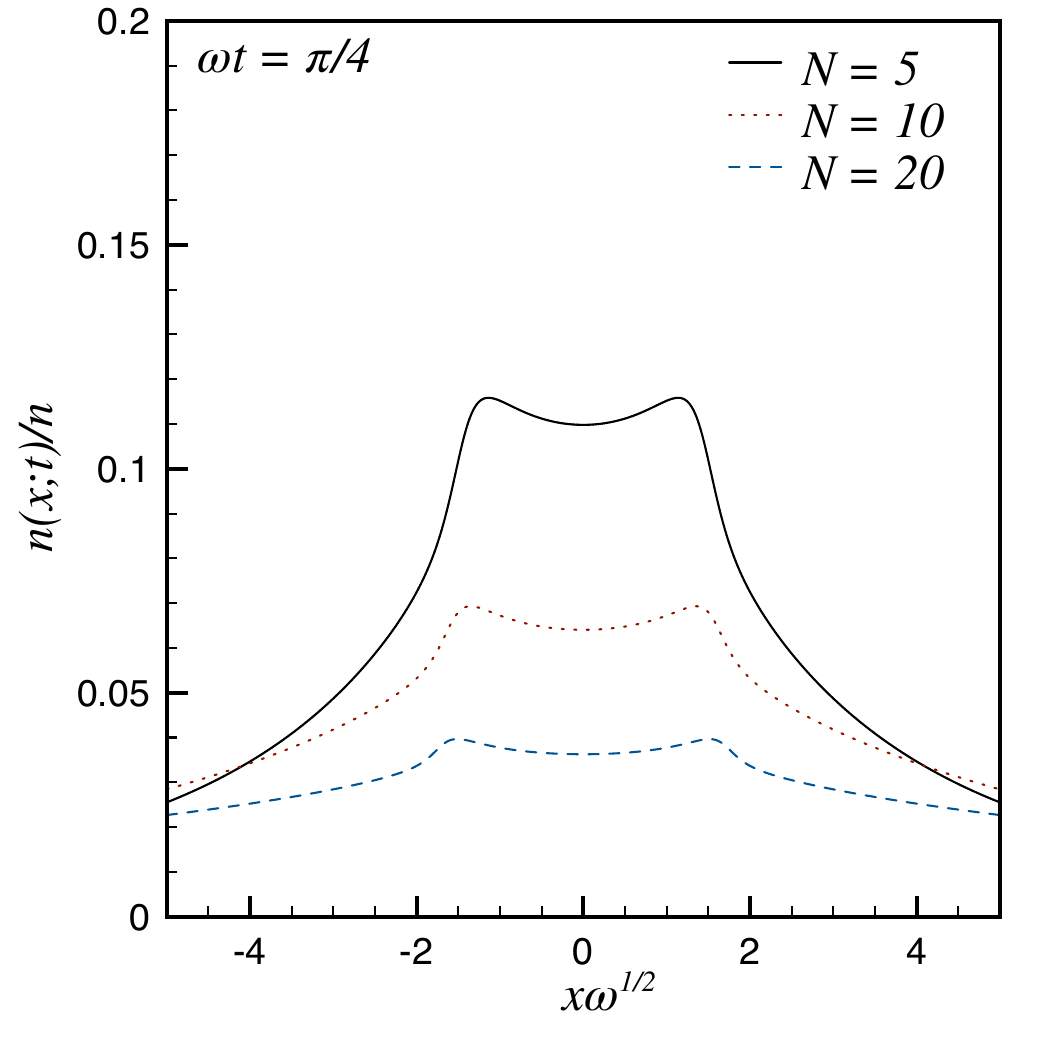}\\
\includegraphics[width=0.33\textwidth]{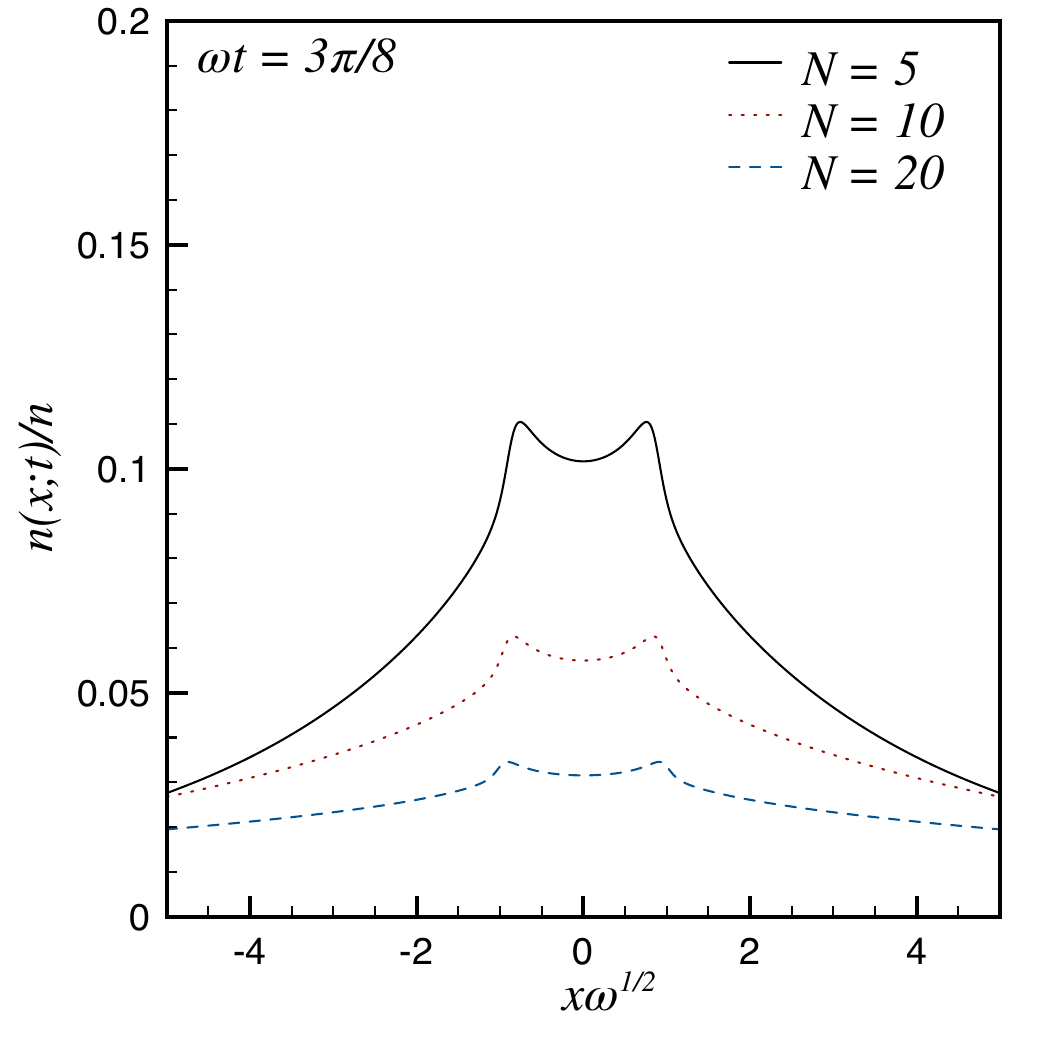}
\includegraphics[width=0.33\textwidth]{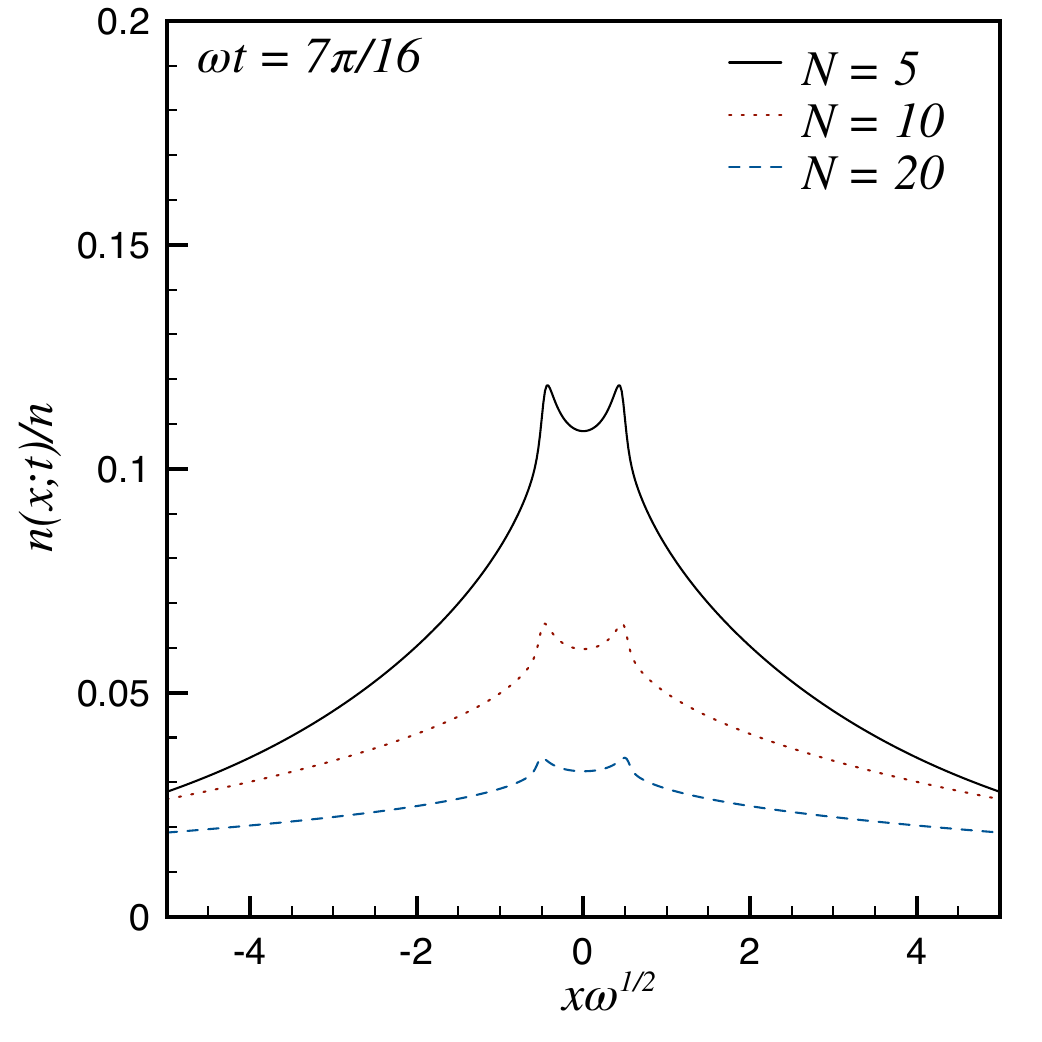}\includegraphics[width=0.33\textwidth]{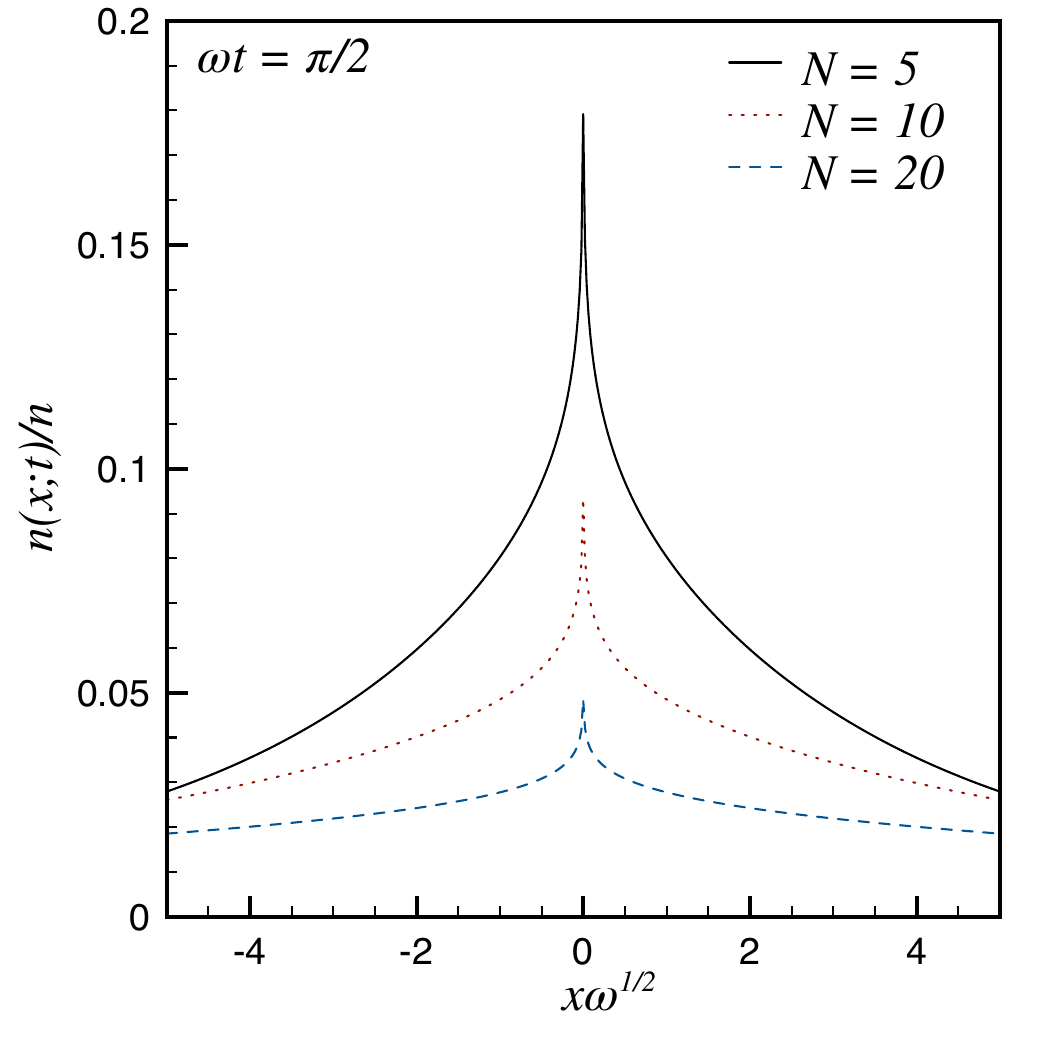}
\caption{Snapshots of the rescaled density $n(x;t)/n$ at different 
rescaled times $\omega t$ as a function of the rescaled distance from the center $x\sqrt{\omega}$.
Different lines represent different initial particle numbers $N$. At time $t=0$ all curves collapse on the initial Gaussian profile. 
} 
\label{fig_nx_t}
\end{figure*}

\subsection{Evolution of the density profile}

The time evolved density is given by Eq. \erf{PsiPsi_t_TDL} 
evaluated at coincident points $x=y,$ i.e.
\begin{multline}
\label{eq_nxt}
n(x;t)  =\\=  \frac{\sqrt{\omega}}{2\pi|\sin(\w t)|}
 \int_{-\infty}^{\infty}dr_0
\frac1
{1+\left(\frac{ \sqrt{\pi}(r- r_0\cos(\w t))}{2N \sin(\omega t)}{\rm e}^{r_0^2}\right)^2},
\end{multline}
where we recall that 
$r=\sqrt{\w} x$. 
It is cumbersome but elementary to show that for $t\to0$, the density profile reproduces the initial Gaussian distribution \erf{eq_nx0}.
The latter is a function of $r=\sqrt{\omega} x$ and $n,$ i.e. it obeys bosonic scaling in the trap. 
Instead at finite times $0<\w t<\pi$, the density is a function of $r,$ $\w t,$ and $N$. 
This shows, as anticipated in the introduction, that a non-equilibrium TDL  in which we keep the density constant and 
eliminate the $N$ dependence cannot be consistently taken at arbitrary time. 
This awkward behavior is a consequence of the different scaling properties of bosons and fermions in the harmonic trap.

 In Fig. \ref{fig_nxt_color} we show the space-time contour plot of $n(x;t)$ for different number of particles $N$.
The dynamics is periodic with period $T=\pi/\omega.$ Moreover, $n(x;t)=n(x,\pi/\w-t)$ holds, so the time evolution between $0<t<T/2$ is reversed between $T/2<t<T$.
The physics is not very transparent from these color plots but it becomes clearer looking at
fixed time slices as those shown in Fig. \ref{fig_nx_t}. 
Let us describe the panels of this figure. 
In the very early stage of the evolution, the cloud expands very quickly with particles moving from the center to the edges of the trap.
During this expansion, the initial Gaussian profile becomes a density profile with slowly decaying tails. 
The expansion is faster for higher number of particles $N$: 
the Pauli principle makes the particles repelling each other after the quench and the effect is larger for tighter initial confinement. 
The {\it undertow} following the expansion creates a local minimum or ``dip'' at the center of the trap, as an unexpected new effect. 
Interestingly, at half-period $\w t=\pi/2$ we observe a ``recondensation'' or ``refocusing'' phenomenon 
(i.e. the density in the center becomes higher), but always in the presence of very fat tails. 
The effect however becomes fainter with increasing number of particles.

 \begin{figure}[t]
\includegraphics[width=0.45\textwidth]{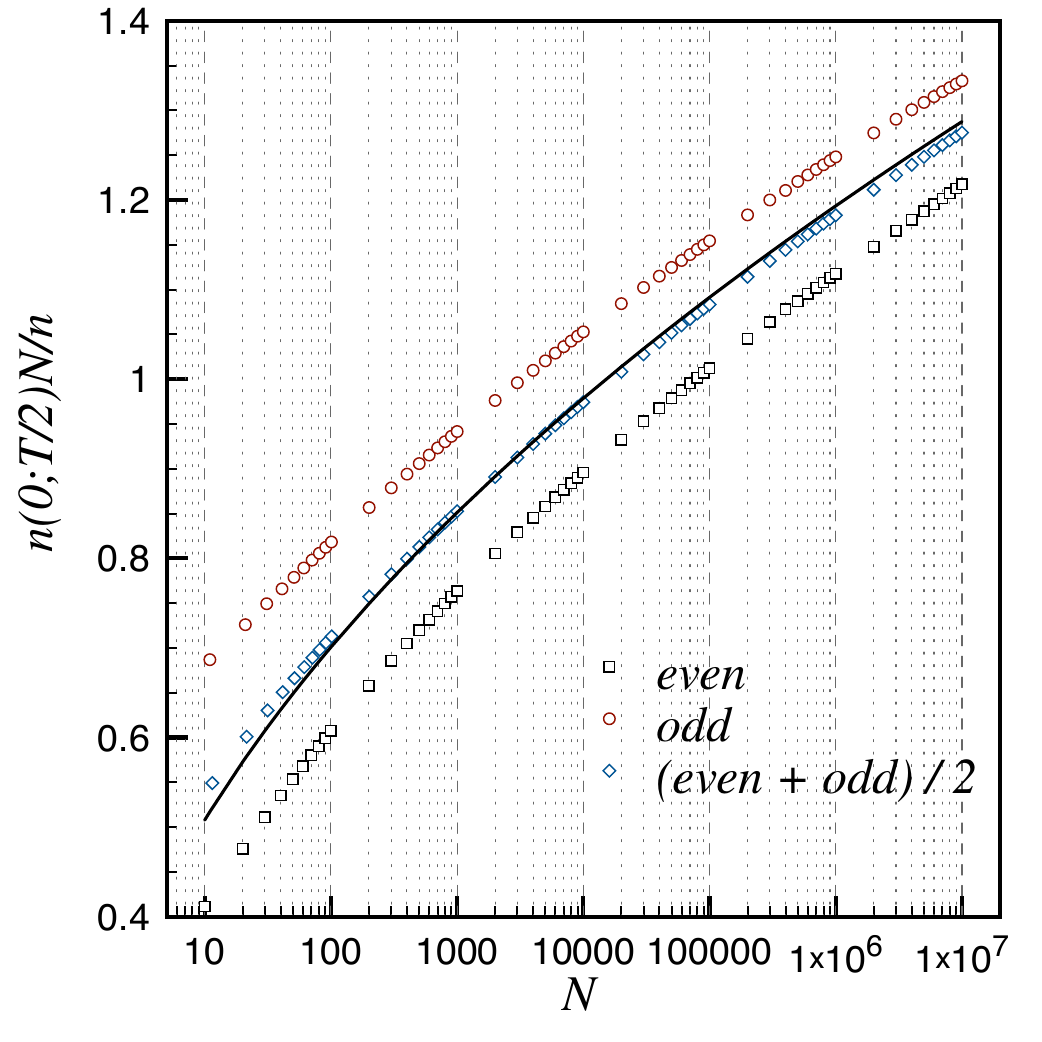}
\caption{Particle density at the center of the trap at time $t=\pi/(2\w)=T/2.$ The numerically evaluated integral (\ref{PsiPsi_0_v2}) using the exact initial correlator \erf{fermi_corr_t_green}
is shown in symbols. The red circles and black squares correspond to $N$ even and odd, respectively, while the blue diamonds represent the average over $N$ and $N+1.$ The asymptotic behavior  (\ref{n0_largeN}) is plotted in solid black line.}
\label{fig_n0_vs_N}
\end{figure}

The refocused density profile at the special time $t=T/2=\pi/(2\w)$ can be obtained analytically as
\begin{multline}
\label{nT/2}
n(x;T/2) 
 =  \frac{\sqrt\w }{2\pi} \int_{-\infty}^{\infty}dr_0
\frac1{1+\left(\frac{ \sqrt{\pi}r}{2N}{\rm e}^{r_0^2}\right)^2} \\
 =  -\frac{\sqrt\w }{2\sqrt{2\pi}}{\rm Li}_{1/2}\left(- \frac{4 N^2}{\pi r^2} \right) \, .
\end{multline}
As was discussed earlier, the apparent divergence at $r = 0$ is an artefact of the TDL in the initial condition, i.e. in replacing 
Eq. \erf{PsiPsi_0_v2} with Eq. \erf{PsiPsi_0_TDL}. For finite but large $N$ the density at $t=T/2$ at the origin behaves as
\be\label{n0_largeN}
n(0;T/2)\approx \frac{n}N \frac1\pi\sqrt{\ln (4N/\sqrt\pi)}.
\ee
In Fig. \ref{fig_n0_vs_N} we compare Eq. (\ref{n0_largeN}) (solid line) with the numerically evaluated density using the exact initial correlation function \erf{fermi_corr_t_green} in the integral (\ref{PsiPsi_0_v2}). 
The exact numerical result shows an oscillation with the parity of $N$ which very slowly disappears in the thermodynamic limit. 
The average of the values for neighboring even and odd $N$ agrees very well with the analytic prediction \erf{n0_largeN}.

\begin{figure}[t]
\includegraphics[width=0.45\textwidth]{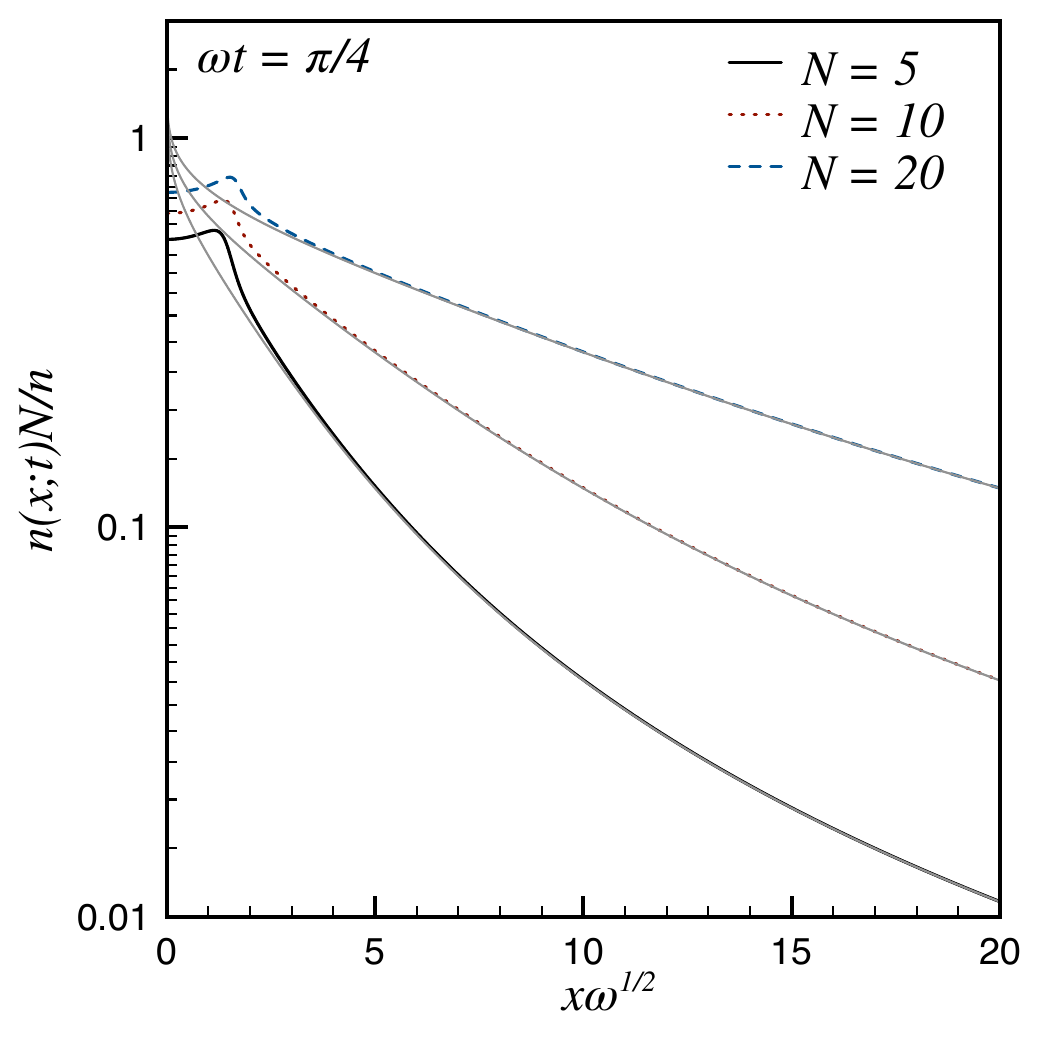}
\caption{Large distance behavior of the rescaled density profile $n(x;t) N/n$ at fixed initial density $n$
and different particle numbers $N.$ The numerically evaluated integral in Eq. (\ref{eq_nxt}) 
is compared with the asymptotic expansion (\ref{nxt_asympt}) (grey lines).}
\label{fig_nxt_scaling}
\end{figure}

Finally, we consider the tails of the density profile by analyzing the large $x$ behavior of \erf{eq_nxt}. 
The tails for $r\gg r^*=\sqrt{\ln(2N\sin\w t/\sqrt\pi)}$
follow from Eq. \erf{Casympt1} by setting $x=y$ (i.e $z=\zeta=0$),  which yields
\be
\label{nxt_asympt}
\frac{n(x;t)}{\sqrt\w}
 \approx    
 -\frac1{2\sqrt{2\pi}|\sin(\omega t)|} {\rm Li}_{1/2}\left(- \frac{4 N^2 \sin^2(\omega t)}{\pi r^2} \right)\, .
\ee
The ${\rm Li}_{1/2}$ function crosses over between two different regimes depending on whether the argument is much larger or much 
smaller than $1$.
Consequently, for any finite value of $N$ there is an intermediate region $r^*\ll r\ll N$ in which \cite{foot1}
\be 
n(x;t)\sim \frac{\sqrt\w}{\pi |\sin(\w t)|}\sqrt{\ln \frac{2N\sin(\w t)}{\sqrt\pi r}}, \quad {\rm for} \; r^*\ll r\ll N,
\ee
which crosses over to  
\be
n(x;t)\sim   \frac{\sqrt{2\w}}{\pi^{3/2}} |\sin(\w t)|\frac{N^2}{r^{2}}, \quad {\rm for} \quad r\gg N > r^*.
\ee
To show the correctness of  the analytic result (\ref{nxt_asympt}), in Fig. \ref{fig_nxt_scaling} we compare it with the large distance behavior of 
the density profile obtained by numerically evaluating the integral (\ref{eq_nxt}) for $\w t=\pi/4$, chosen as a representative time.

 \subsection{Time averaged density profile}

The time average over a period of  Eq. \erf{C_GGE_def} is 
\begin{multline}
\overline{n(x;t)}  \equiv  \frac1T \int_0^T dt \,n(x;t)  \\
=\frac{\sqrt{\omega}}{2\pi^2}
 \int_{-\infty}^{\infty}dr_0
 \int_0^\pi \frac{d\tau}{\sin\tau}
\frac1{1+\left(\frac{ \sqrt{\pi}(r- r_0\cos(\tau))}{2N \sin(\tau)}{\rm e}^{r_0^2}\right)^2}\\
= \frac{\sqrt{\omega}r }{2\pi^2} \int_{-\infty}^{\infty}db \int_{-1}^1 ds \frac1{1-s^2+a(1-bs)^2},
\end{multline}
where $\tau=\w t$, $s=\cos\tau$, $b=r_0/r$, and we introduced
\be
a=\frac{\pi r^2}{4N^2}{\rm e}^{2r_0^2}.
\ee
The $s$-integral can be worked out yielding 
\be
\label{navres}
\overline{n(x;t)} =\frac{\sqrt{\omega} r}{2\pi^2} \int_{-\infty}^{\infty}d b
\frac{\ln\frac{s_1-1}{s_1+1}-\ln\frac{s_2-1}{s_2+1}}{(ab^2-1)(s_1-s_2)}\,,
\ee
where $s_{1,2}$ are the roots of the equation $(ab^2-1)s^2-2ab\,s+(a+1)=0.$

Eq. \erf{navres} is valid for arbitrary $x$. Its large distance behavior is readily obtained using Eq. (\ref{nxt_asympt}) 
that, changing integration variable to $s=\cos\w t,$ leads to 
\begin{multline}
\label{navres_large}
\overline{n(x;t)}/\sqrt\w  
 \approx  -\frac{1}{(2\pi)^{3/2}} \int_{-1}^{1}ds 
\frac{{\rm Li}_{1/2}\left(- \frac{4 N^2}{\pi r^2} (1-s^2) \right)}{1-s^2} \\
 =  -\frac{\sqrt{2}}{4\pi} \sum_{k=1}^{\infty} \frac{(k-1)!}{\sqrt{k}(k-1/2)!} 
\left( - \frac{4 N^2}{\pi r^2} \right)^{k} .
\end{multline}

\begin{figure}[t]
\includegraphics[width=0.45\textwidth]{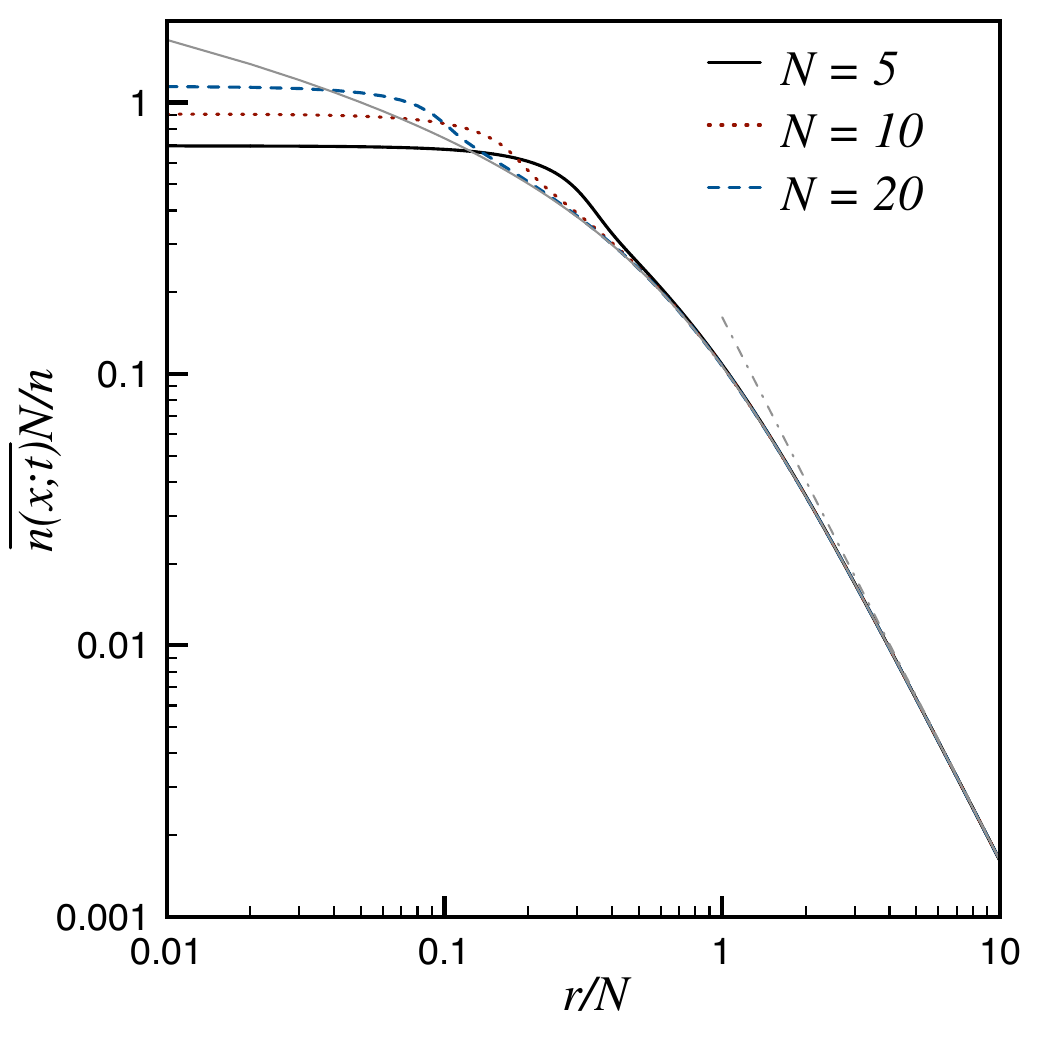}
\caption{Large distance behavior of the rescaled time-averaged density profile
$\overline{n(x;t)} N/n$ against $r/N$ for different particle numbers on a double logarithmic scale. 
The asymptotic is captured by Eq. (\ref{navres_large})  (grey thin solid line)
which for $r/N \to \infty$ decays as $2\sqrt{2}/\pi^{5/2} (r/N)^{-2}$ (grey thin dot-dashed line).
} 
\label{fig_nxt_avg_scaling}
\end{figure}

In Fig. \ref{fig_nxt_avg_scaling} we report the time average (\ref{navres}) for different $N$ showing that, for sufficiently large distances, 
all curves collapse on the same function of the rescaled variable $r/N$ given by Eq. \erf{navres_large}.

\section{The generalized Gibbs ensemble}\label{sec:stat}

One of the main results about the non-equilibrium quench dynamics of translational invariant integrable systems is that 
local observables attain stationary values which are the same as if the entire system was in a statistical ensemble in which all the relevant 
integrals of motion are taken into account as dynamical constraints. 
This statistical ensemble is the Generalized Gibbs Ensemble (GGE) \cite{gg} 
in which the conservation of a complete set of local and quasi-local integrals of motion is imposed \cite{cazalilla-2006,barthel-2008,cramer-2008,cramer-2010,calabrese-2011,calabrese-2012,cic-12,fagotti-2013,p-13,fe-13b,sc-14,fcec-14,wdbf-14,PMWK14,KBC:Ising,ilievski-2015a,pvc-16,pvw-17,vidmar-2016,ef-16}.

In the case of non-homogeneous systems, the literature is much less clear. As the role of locality is less stringent, 
it is natural to wonder whether a GGE may describe some relevant features of the system. 
In the case studied in this manuscript, stationary values are not attained for long times so we can only ask whether a GGE captures
the time averaged values of some local observables \cite{rigol-06}.
The most reasonable set of conserved charges to be considered here are the fermionic mode occupation numbers 
$\hat n_q$ in Eq. \erf{Hwithnq}. 

Consequently, the goal of this section is to study some properties of  the GGE 
\be
\hat\rho_\text{GGE}= Z^{-1} \exp\Big(-\sum_q \lambda_q \hat n_q\Big), 
\ee
where $Z$ ensures that $\mathrm{Tr}\,\hat\rho_\text{GGE}=1$. 
The Lagrange multipliers $\lambda_q$ are fixed by the requirement that each integral of motion 
$\hat n_q$ assumes the same value in the initial state ($\langle \hat n_q \rangle\equiv \langle \psi_0|\hat n_q|\psi_0 \rangle$)
and in the GGE ($ \langle \hat n_q\rangle_{\rm GGE}\equiv{\rm Tr} [\hat\rho_\text{GGE}\,\hat n_q]$).
For a free-fermionic model, solving the GGE conditions $\langle \hat n_q \rangle= \langle \hat n_q\rangle_{\rm GGE}$
is straightforward since the mode occupation satisfies 
Fermi statistics and hence $ \langle \hat n_q\rangle_{\rm GGE}= (1+e^{\lambda_q})^{-1}$ leading to 
$e^{\lambda_q}=1/\langle \hat n_q \rangle-1$. 

The elementary building blocks of the GGE are then the expectation values of the fermionic mode occupations in the initial state 
$\langle \hat n_q\rangle$. 
These can be obtained from the initial correlator  \erf{PsiPsi_0_v2} of the real-space fermionic fields 
$\langle\hat\Psi^{\dag}(x)\hat\Psi(y)\rangle.$
From the knowledge of $\langle \hat n_q\rangle$  we can, in principle, recover the expectation value in the GGE of an arbitrary {\it local} observable
and of its correlations at finite distance. 
In the next subsection we explicitly work out $\langle \hat n_q\rangle$ for large $N$ and we show that from the resulting GGE we may 
recover the time-average density profile \erf{navres}. 

The time average of Eq. (\ref{fermi_corr_t_eta}) is given by
\be
\label{C_GGE_def}
 \overline{C(x,y;t)}   =  \sum_{q=0}^\infty \varphi^*_q(x)\varphi_q(y) \langle\hat n_q \rangle,
\ee
which only depends on the fermionic mode occupations. 
Given that by GGE construction $\langle\hat n_q \rangle=\langle\hat n_q \rangle_{\rm GGE}$ and that 
$\langle\hat \eta^\dag_q\eta_p \rangle_{\rm GGE}=\delta_{qp} \hat n_q$, the time-averaged fermionic correlation 
and the GGE one $C_{\rm GGE}(x,y)\equiv {\rm Tr} [\rho_{\rm GGE} \Psi^\dag (x) \Psi(y)]$ are equal.
This implies that also the density profiles in the GGE and the time-averaged one are equal, since they are just the 
corresponding correlations for $x=y$.

 \subsection{Fermionic mode occupation}
The expectation values of the conserved fermionic mode occupations $\langle\hat n_{q}\rangle$ are obtained by plugging
the initial correlation function (\ref{PsiPsi_0_v2}) in the definition of the fermionic modes (\ref{psi_eta}):
\be
\label{nq_int1}
\langle \hat n_q \rangle  =\int_{-\infty}^{\infty} dx \int_{\infty}^{\infty} dy \, \varphi_q(x)\varphi^{*}_q(y)
\langle\hat\Psi^{\dag}(x)\hat\Psi(y)\rangle .
\ee
In order to proceed we plug in this expression the TDL (\ref{PsiPsi_0_TDL}) of $\langle\hat\Psi^{\dag}(x)\hat\Psi(y)\rangle$. 
To ensure that this approximation does not introduce systematic errors, 
we evaluated the double integral \erf{nq_int1} numerically using both the TDL (cf. \erf{PsiPsi_0_TDL}) and the finite $N$ (cf. \erf{PsiPsi_0_v2}) 
expressions of the initial correlation function and found that for sufficiently large $N$ they lead to the same result.

Thus the desired mode occupation can be written as 
\begin{multline}
\label{nq_int2}
 \langle \hat n_q \rangle 
  =  \int_{-\infty}^{\infty} dx \int_{\infty}^{\infty} dy \, \varphi_q(x)\varphi^{*}_q(y) \\
 \times \frac{n}{\sqrt{\pi}} \,{\rm e}^{-r^2} \exp \left( -\frac{2 n}{\sqrt{\pi}} {\rm e}^{-r^2} |z| \right),
\end{multline}
where $z=x-y$ and $r=\sqrt\w(x+y)/2$.
As long as we keep $q$ finite, in the TDL ($N\to\infty,$ $\w\to0$ with $N\sqrt\w$ fixed) the previous integral
gives $\langle \hat n_q \rangle \to 1$ for all $q.$ 
However, for $N$ large but finite, whenever $q \sim  \ln N$, the mode occupation $\langle \hat n_q \rangle$ 
deviates from one (see the Appendix).

In Fig. \ref{fig_nq} we report the rescaled mode occupation $N\langle \hat n_{q} \rangle$ as a function of $q/N^2$.
This figure provides a strong numerical evidence that the data for all values of $N$ collapse on a universal smooth function. 
Therefore, in the TDL, Eq. (\ref{nq_int2}) for the modes with $q/N^2\sim O(1)$  greatly simplifies.
Indeed,  we can use the asymptotic expansion of the Hermite polynomials for large $q$
\be
\label{asympt}
{\rm e}^{-\omega x^2/2} {\rm H}_{q}(x\sqrt{\omega}) \sim \frac{2\Gamma(q)}{\Gamma(q/2)} \cos (x\sqrt{2\omega q} - q\pi/2),
\ee
and for large $q$ and large $N$ (with $N\sqrt{\omega} = n$ fixed)
 we can approximate Eq. (\ref{nq_int1}) as
\begin{multline}
\label{nqstart}
\langle \hat n_q \rangle  \approx  \frac{n}{\pi} \sqrt{\frac{2}{\pi q}} \int_{-\infty}^{\infty} dr  \int_{-\infty}^{\infty} dz \\
   \times [\cos(z \sqrt{2\omega q})+(-1)^{q}\cos(2 r \sqrt{2 q}) ] \\
   \times {\rm e}^{-r^2} \exp \left( -\frac{2 n}{\sqrt{\pi}} {\rm e}^{-r^2} |z| \right),
\end{multline}
 where we used 
 $(2\Gamma[q]/\Gamma[q/2])^2 / (2^q q!) \approx \sqrt{2/(\pi q)}$ for $q\gg 1$.
For $\omega\to0$ keeping $\omega q$ constant, the term proportional to $(-1)^{q}$ is highly oscillating
and vanishes in the TDL. 
The smooth term is easily integrated over $z$ leading to
 \be
\label{nq_TDL}
\begin{split}
\langle \hat n_q \rangle & \approx  \frac{1}{\pi\sqrt{2\,q}}\int_{-\infty}^{\infty} dr  
\frac{ {\rm e}^{-2r^2} }{{\rm e}^{-2r^2} + \pi q/(2N^2)} \\
&= \frac{1}{N}  \left[ -\frac{1}{2\sqrt{2\, \tilde q}} \, {\rm Li}_{\frac{1}{2}}\left(-\frac{1}{\tilde q} \right) \right]
\equiv  \tilde n (\tilde q),
\end{split}
\ee
where  $\tilde q\equiv \pi\omega q/(2n^2) = \pi q /(2N^2)$. 
This analytic result is compared in Fig. \ref{fig_nq} with the numerical evaluation of the mode occupation 
showing an excellent agreement for large enough $N$. Moreover, as a consequence of 
$-\int_{0}^{\infty}\frac{dz}{\pi\sqrt{2\, z}} \, {\rm Li}_{\frac{1}{2}}\left(-\frac{1}{z} \right) =1,$ 
Eq. (\ref{nq_TDL}) satisfies the sum rule $\sum_{q} \langle \hat n_q \rangle = N,$
suggesting that, in the TDL, only a vanishing fraction of the particles are not in modes with $q\sim O(N^2)$, 
see the Appendix for more details.

 \begin{figure}[t]
\includegraphics[width=0.45\textwidth]{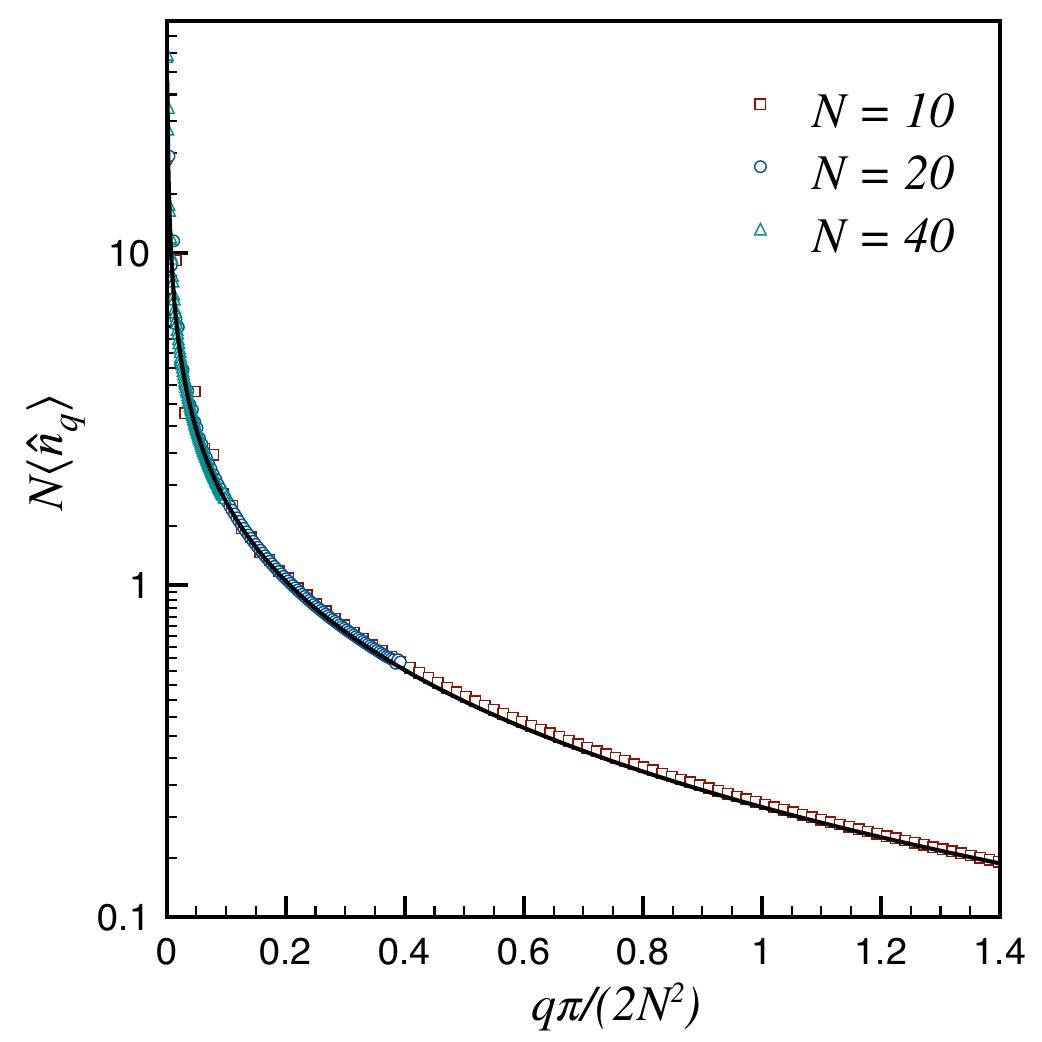}
\caption{Rescaled fermionic mode occupation $N \langle \hat n_{q} \rangle$ 
in logarithmic scale as function of $ q\pi/(2N^2)$ for different particle numbers $N$. 
The numerical data, evaluated using Eq. (\ref{nq_int1}), collapse on the asymptotic universal function (full black line)
given by Eq. (\ref{nq_TDL}). 
} 
\label{fig_nq}
\end{figure}

The asymptotic behavior of Eq. \erf{nq_TDL} for both small and large values of $\tilde q$ is easily worked out.
For $\tilde q \gg 1$ the distribution $\langle \hat n_q \rangle$ shows a power law decay
$N\langle \hat n_q \rangle  \simeq 1/(2\sqrt{2} \tilde q^{3/2})$. 
For $q\ll1$, we use the asymptotic expansion of the Polylogarithm function for $-z\gg 1$ \cite{foot1}
to obtain $N\langle \hat n_q \rangle  \simeq \sqrt{-\ln(\tilde q) / (2 \pi \tilde q)}$.
This divergence as $\tilde q\to 0$ is a consequence of the rescaling because, as already stressed, as $q\to0$ we have 
$\langle \hat n_q \rangle\to1$.

 \begin{figure}[t!]
\includegraphics[width=0.45\textwidth]{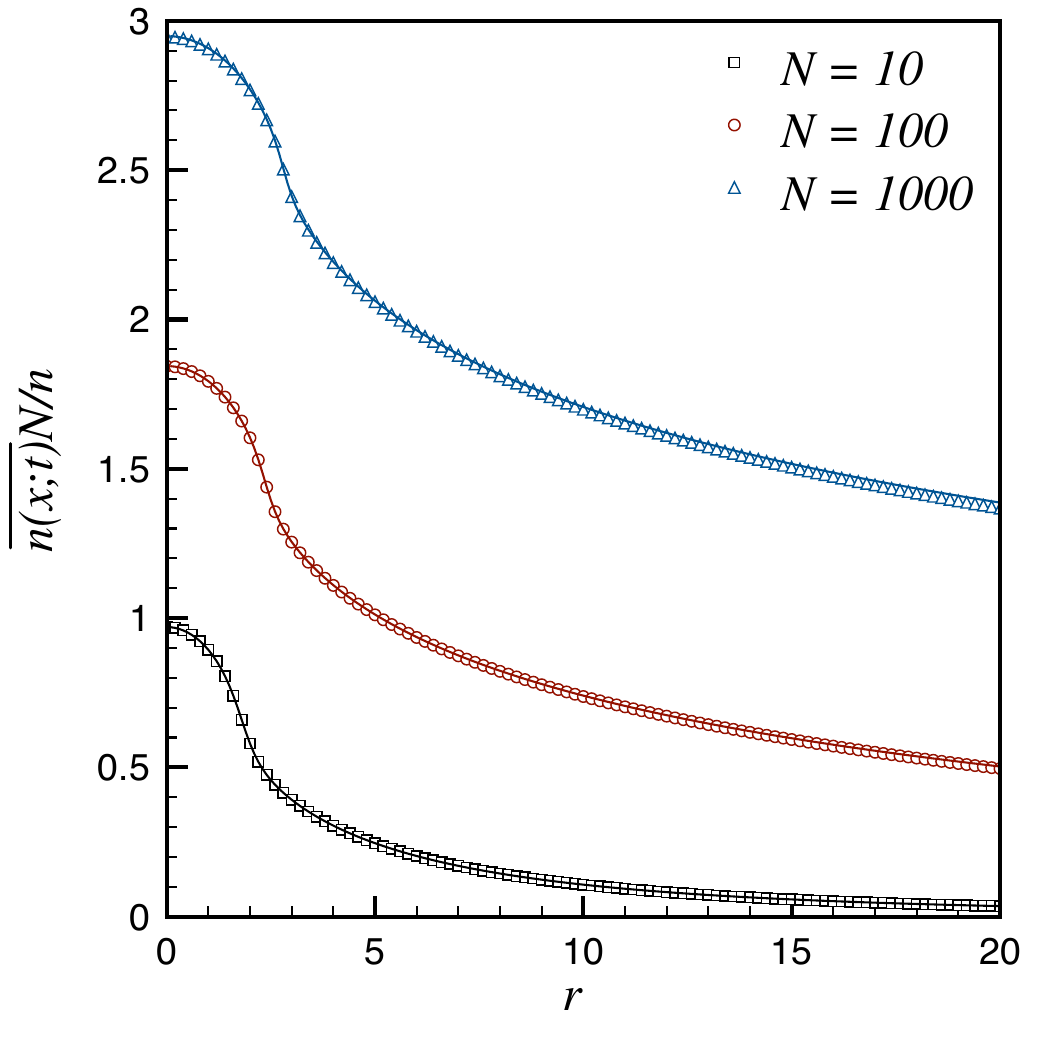}
\caption{Comparison between the time-averaged density $\overline{n(x;t)}$ (\ref{navres})
(full lines) and the GGE density $n_\text{GGE}(x)$  \erf{n_eq} (symbols)
vs. $r=\sqrt{\w} x$ for different particle numbers $N$.
In Eq. \erf{n_eq} we used the conservative values of $q_1=20$ and $q_2=1000$ for the three values of $N$ reported. 
} 
\label{fig_nxt_avg_compare}
\end{figure}

\subsection{The density profile in the GGE for finite $N$}

The goal of this subsection is to show that it is possible to obtain directly the GGE density profile from $\langle \hat n_q \rangle$ 
and to numerically recover, for finite but large $N$, the time-average \erf{navres} as it should be by construction.

The GGE density profile $n_\text{GGE}(x)$ in terms of the mode occupations is simply
\be
n_\text{GGE}(x)= \sum_{q=0}^{\infty}  |\varphi_{q}(x)|^2 {\rm Tr} [\hat n_q \hat\rho_{\rm GGE} ]=
\sum_{q=0}^{\infty} \langle \hat n_q \rangle |\varphi_{q}(x)|^2\,.
\ee 
The asymptotic scaling form (\ref{nq_TDL}) is  valid for large $q,$ so we split $n_\text{GGE}(x)$ in two sums
\be
n_\text{GGE}(x) = \sum_{q=0}^{q_1} \langle \hat n_q \rangle |\varphi_{q}(x)|^2 
+ \sum_{q=q_1+1}^{\infty} \tilde n(\tilde q) |\varphi_{q}(x)|^2 \, ,
\label{ngge_q1}
\ee 
where $q_1 \sim \ln(N).$  For $q<q_1$ we have $ \langle \hat n_q \rangle\simeq 1$.
At this point the second sum can be approximated by an integral and the above equation should reproduce \erf{navres}.
Anyhow, the situation is not so simple because there are severe finite-size effects on $ \langle \hat n_q \rangle$ as numerically 
shown in the appendix. Thus Eq. \erf{ngge_q1}, although correct, is accurate only for enormous values of $N$.

In order to have a prediction for moderate large values of $N$ we heuristically proceed as follows. 
We further split into two parts the second sum in \erf{ngge_q1}:  
for $q\gtrsim q_2\approx 2r^2$ we use Eq.  \erf{asympt} to approximate the wave function and we analytically work out the integral 
in which the rapidly oscillating $\cos^2(\dots)$ factor is replaced by $1/2$.
Then for the sum up to $q_1$, we do not approximate  $ \langle \hat n_q \rangle\simeq 1$, but we use 
the actual value of  $ \langle \hat n_q \rangle$ which for finite $N$ is slightly different from one.  
Putting together the three pieces constructed in this was we thus finally obtain
\be
\label{n_eq}
\begin{split}
n_\text{GGE}(x) &= \sum_{q=0}^{q_1} \langle \hat n_q \rangle |\varphi_{q}(x)|^2 
+ \sum_{q=q_1+1}^{q_2} \tilde n(\tilde q) |\varphi_{q}(x)|^2 \\
 &-  \frac{{\rm Li}_{3/2} (-1/\tilde q_2)}{(2\pi)^{3/2}} \, ,
\end{split}
\ee
where  $\tilde q_2 =  \pi (q_2+1)/(2N^2)$. 
This result is compared with the time-averaged profile \erf{navres} in Fig. \ref{fig_nxt_avg_compare}:
in spite of the heuristic reasoning the agreement is perfect. 
In the spirit of the GGE, this shows that it is possible to obtain time-averaged values without solving the 
entire many-body dynamics, but only from the knowledge of the integrals of motion in the initial state.

\section{Conclusion}
\label{sec:concl}

We studied the non-equilibrium dynamics after a quantum quench of a one dimensional Bose gas in the presence of an external trapping 
harmonic potential. We consider  the case in which the interaction parameter is quenched from zero to infinity, i.e. we switch  
from non-interacting to strongly interacting bosons. 
The setup breaks translational invariance so the particle density and the two-point fermionic correlation function
have a nontrivial time evolution, as a first important difference compared to the homogeneous case \cite{kcc14}.

The main physical and technical difficulty in this study is related to the nature of the thermodynamic limit. 
Indeed, in the initial state all bosons are in the same one-particle state, and 
the trap frequency $\w$  introduces a natural length scale $1/\sqrt\w$ for the bosonic thermodynamic quantities which is independent 
of the number of particles. 
However, the time evolution is governed by an essentially fermionic Hamiltonian, 
and the Pauli principle favors a different typical length scale of the thermodynamic quantities which is $\sqrt{N/\w}$
and does depend on the number of particles. 
Here we overcame this difficulty, but the price to pay is that various observables, lengths, distances and all physical variables 
have to be rescaled in a non-standard way with the number of particles $N$.

Because of the strange interplay between the bosonic and fermion nature of the problem, there are several novel and unexpected 
features which have no counterpart in the homogeneous case \cite{kcc14}. 
First, the non-equilibrium dynamics is exactly periodic with period $T=\pi/\w$ and relaxation never occurs. 
This ``breathing'' of the particle cloud in the trap is not a surprise, it is a consequence of having equally spaced single particle levels 
and it is common to many non-equilibrium situations of one-dimensional Bose gas in a harmonic trap. 
However, it is a marked difference with the homogeneous problem \cite{kcc14}, as well as with the evolution in a hard-wall trap
\cite{mckc2014}, where relaxation dynamics is always the rule.

As soon as the  the infinitely repulsive interaction is switched on, the cloud of bosons starts expanding because of the 
strong repulsion. This expansion becomes faster as the particle number increases as a consequence of the tighter confinement.
To characterize this dynamics, we provided analytic results for the equal-time fermionic  two-point correlation function. 
Although not being experimentally measurable,  this is the building block for the calculation of all observables. 
We used this correlation to obtain an analytic expression for the particle density valid for large number of particles: 
remarkably, the originally localized particle cloud with Gaussian decaying tails acquires a much broader shape characterized by 
an algebraic large distance decay.
The expansion is so fast that there is a undertow effect such that for certain times the density has a 
pronounced local minimum at the center of the trap.  
At half period $t=\pi/(2\w)$ we observe a refocusing phenomenon with the density showing again a sharp peak at the center, 
but always with algebraically decaying tails at large distance. 
Interestingly, while in the initial state the average density is $n=N \sqrt\w$, at finite time as well as for the time-average value we have 
densities which are proportional to $n/N$ which is suppressed even with respect to the density of the fermions 
ground state $n^f= n/\sqrt N$.
A physical explanation of this behavior could be that the initial confinement is so tight that the particles have 
the energy to move much further away compared to the equilibrium configuration.

We also worked out the GGE built with the conserved fermionic mode occupations.
The latter have been analytically obtained in the large particle number limit, showing that they obey a nontrivial scaling for high energy modes.
We showed that the density distribution computed in this GGE coincides, for large $N$, with the time-averaged profile.

Many relevant observables have not been computed here and require further investigations
Among these the most important one is probably the bosonic two-point correlation which has been obtained in the stationary state both 
for the homogeneous case \cite{kcc14} and for hard-wall confinement \cite{mckc2014}. 
Its time evolution required the use of more elaborated techniques and has been possible only for homogeneous systems \cite{dc-14,pc-17}.
An analogous calculation in the harmonic trap seems much harder than in the above cases, even if limited to the time-average expectation. 
Another open issue concerns the time evolution of the entanglement entropy.
For the homogeneous case, the stationary state entanglement is known \cite{ckc-14} and, by using the quasi-particle spreading of 
entanglement \cite{cc-05}, it has been possible to work out the entire time evolution \cite{ac-16}.
It is a remarkable problem to understand how the quasi-particle spreading is modified in this non-homogeneous quench 
also in view of a recent cold-atom experiment \cite{kaufman-2016}.
Most of the known approaches for the entanglement entropy in inhomogeneous settings only deal with the low-energy physics 
(as e.g. in \cite{sc-08,dsvc17}), while in a quench setup high-energy modes are governing the dynamics.

\begin{acknowledgments}
We thank Brayden Ware for helping at choosing the appropriate word `undertow'.
M.C. acknowledges support by the Marie Sklodowska-Curie Grant No. 701221 NET4IQ. 
M.K. acknowledges funding a ``Pr\'emium'' Postdoctoral Grant of the Hungarian Academy of Sciences. 
\end{acknowledgments}

\begin{appendix}
\section{Finite-$N$ scaling of $\langle \hat n_q \rangle$}

In this appendix we provide the details about the crossover of $\langle \hat n_q\rangle$ between the two regimes in $q$
which takes place around $q\sim \ln N$.
Our starting point is the exact form of $\langle \hat n_q \rangle $: 
\begin{multline}\label{nq_app1}
 \langle \hat n_q \rangle 
  =  \int_{-\infty}^{\infty} dr \int_{\infty}^{\infty} dz \, A \, {\rm e}^{-\frac\w4 z^2-2 A |z|}  \\
   \times \frac{{\rm e}^{-r^2} }{2^{q} q! \sqrt{\pi}}H_{q}\left(r+\frac{\sqrt\w z}{2}\right) H_{q}\left(r-\frac{\sqrt\w z}{2}\right) \, ,
\end{multline}
where $A= n \exp(-r^2)/\sqrt{\pi}$.
Since we are interested in the large $N$ behavior, we proceed by expanding the Hermite polynomials in power of $\sqrt\w$. 
Using  
\be
\partial^{k}_{x} H_{q}(x) = 2^{k} k! \binom{q}{k} H_{q-m}(x) \, ,
\ee
one  obtains
\begin{multline}
H_{q}\left(r+\frac{\sqrt\w z}{2}\right) H_{q}\left(r-\frac{\sqrt\w z}{2}\right) =\\
 =     \sum_{k,p=0}^{q}(-1)^{k} \binom{q}{k} \binom{q}{p}   
 \times H_{q-k}(r)H_{q-p}(r) (\sqrt\w z)^{k+p},
\end{multline}
where in the sum only terms with $k+p$ even contribute.
Plugging the previous result in Eq. (\ref{nq_app1}), the integral over $z$ can be evaluated as 
\be
\int_{-\infty}^{\infty} dz  z^{p+k} A {\rm e}^{-\frac\w4 z^2-2 A |z|}=
(-4)^\frac{p+k}{2}\partial^\frac{p+k}{2}_{\omega} \mathcal{I}(2A/\sqrt\omega),
\ee
with
$
 \mathcal{I}(z) = \sqrt\pi z \exp(z^2) {\rm erfc}(z),
$
which finally leads to the following expansion 
\begin{multline}
\label{nq_app2}
 \langle \hat n_q \rangle  \simeq 
  \sum_{p,k=0}^{q}  \frac{(-1)^{k} (-4)^\frac{p+k}{2} q!}{p! k! \sqrt{(q-k)!(q-p)!}} 
  \left(\frac{\sqrt\w}{\sqrt{2}}\right)^{k+p}  \\
  \times\int_{-\infty}^{\infty} dr \, \tilde\varphi_{q-k}(r)\tilde\varphi_{q-p}(r) 
 \left[\partial^\frac{p+k}{2}_{\w} \mathcal{I}(2A/\sqrt\w) \right]  \, ,
\end{multline}
where $\tilde\varphi_{k}(r)$ are the harmonic oscillator eigenfunctions with trap frequency $\omega = 1$.

 \begin{figure}[t]
\includegraphics[width=0.45\textwidth]{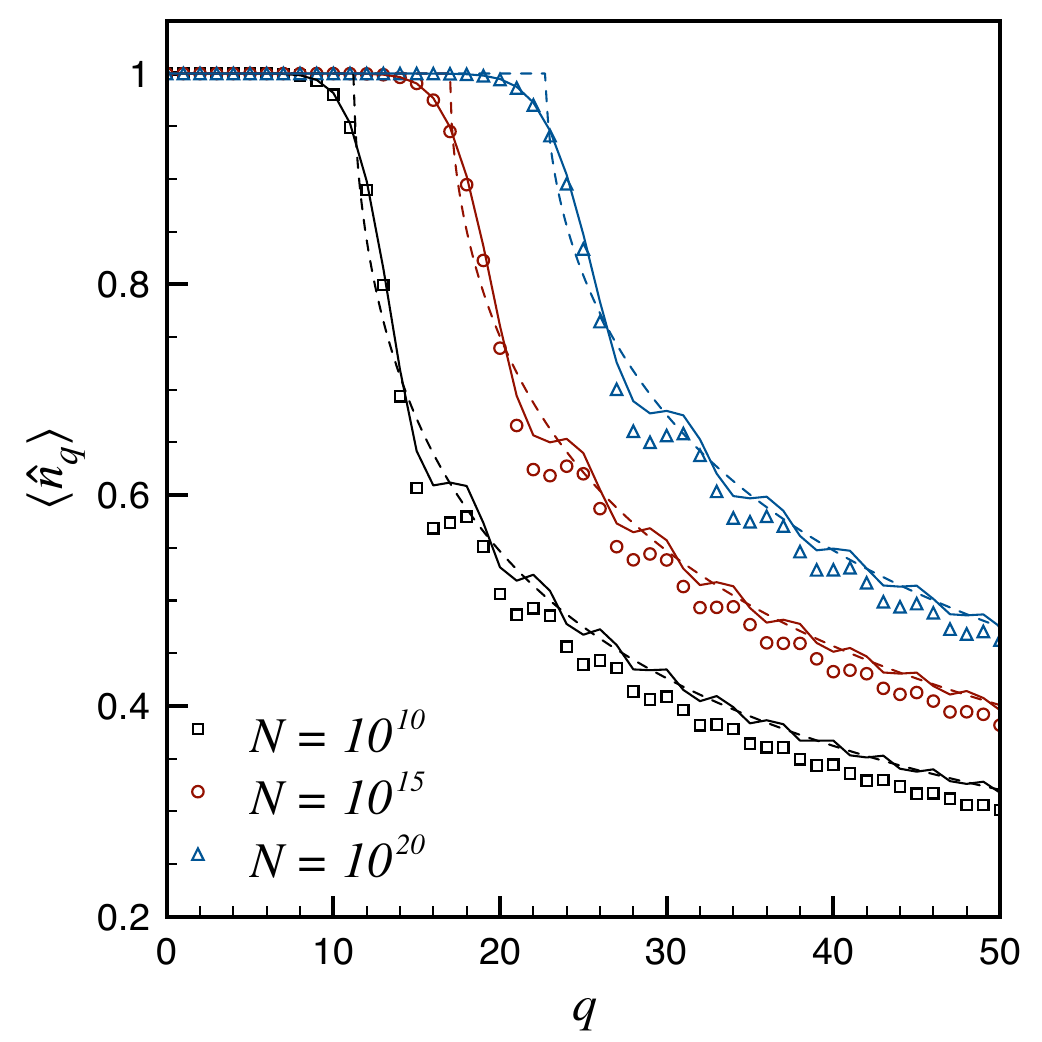}
\caption{ \label{fig:nqlargeN} Mode occupation for very large $N$.
The exact numerically evaluated integral (\ref{nq_int1}) (symbols) is compared with Eq. (\ref{nq_app3}) (full lines) 
and with the semiclassical approximation (\ref{nq_app4}) (dashed lines).
} 
\label{fig_nq_log}
\end{figure}

For $\w \to 0$, the leading term is the one with $p=k=0$. Furthermore we can approximate
$\mathcal{I}(2A/\sqrt\w) \approx \Theta(2 A^2 - \w/4)$,  which is  justified by the Gaussian cut-off
in the $r$-integral. Within this approximation we get 
\be
\label{nq_app3}
 \langle \hat n_q \rangle  \simeq  \int_{-r^{*}}^{r^{*}} dr \, \tilde\varphi_{q}(r)^2 ,
\ee
where $r^{*} = \sqrt{\ln(8N^2/\pi)/2}$ is the positive root of the equation $2A^2-\w/4 = 0$.
In Fig. \ref{fig_nq_log} Eq. \erf{nq_app3} is compared with the exact numerical evaluation of $ \langle \hat n_q \rangle$
finding a reasonable agreement. 

A closed form for  Eq. (\ref{nq_app3}) can be obtaineded in the semiclassical limit.
In this limit, $\tilde\varphi_{q}(r)^2 \simeq (2q+1-r^2)^{-1/2}/\pi$ for $r\in[-\sqrt{2q+1},\sqrt{2q+1}]$.
Therefore, as far as $\sqrt{2q+1}< r^{*}$, i.e. for $q < q^{*} \equiv \ln(8N^2/\pi)/4-1/2 $,
$\langle \hat n_q \rangle = 1$ because of the normalisation condition.  
Conversely, for $q > q^{*}$
\be\label{nq_app4}
\langle \hat n_q \rangle \simeq \frac{2}{\pi} \arctan \left( \frac{r^{*}}{\sqrt{2q+1 - r^{*2}}} \right)\, ,
\ee
which is an universal scaling function in terms of the rescaled variable $q/r^{*2}$.

\end{appendix}



\begin{thebibliography}{99}
\bibitem{foot1}
We use the asymptotic expansion of the Polylogarithm function for $-z\gg 1$, i.e. 
${\rm Li}_{s}(z) = \sum_{k=0}^{\infty} (-1)^{k}(1-2^{1-2k})(2\pi)^{2k}  \frac{B_{2k}}{(2k)!} \frac{\ln(-z)^{s-2k}}{\Gamma[s+1-2k]}$.
\end{thebibliography}
\end{document}